\documentclass[10pt]{article}
\usepackage{graphicx}
\usepackage{wrapfig}
\usepackage{fancyhdr}
\usepackage[ansinew]{inputenc}
\usepackage{color}
\usepackage{dsfont}

\usepackage{amsmath, amssymb, amsfonts, amsthm, bm, natbib} % write standard N, R, Z, Q symbols for number sets.
\usepackage{adjustbox}
\def\ds{\displaystyle}
\graphicspath{{./Pictures/}} % Graphics Path to find your pictures
\begin{document} %%%%%%%
%%%%%%%%%%%%%%%%%%%%%%%%

% Title
\begin{flushleft}
{\Large \bf Identifying down and up-regulated chromosome regions using RNA-Seq data.} %\footnote{Accepted for publication in ... \\ See http://...}}
\end{flushleft}

\vspace{5pt}

{\flushleft 
{\bf Vin\'{i}cius Diniz Mayrink \, and \, Fl\'{a}vio Bambirra Gon\c{c}alves} \\
Departamento de Estat\'{i}stica, Universidade Federal de Minas Gerais, Brazil.}

\vspace{10pt}

\begin{abstract}
The number of studies dealing with RNA-Seq data analysis has experienced a fast increase in the past years making this type of gene expression a strong competitor to the DNA microarrays. This paper proposes a Bayesian model to detect down and up-regulated chromosome regions using RNA-Seq data. The methodology is based on a recent work developed to detect up-regulated regions in the context of microarray data. A hidden Markov model is developed by considering a mixture of Gaussian distributions with ordered means in a way that first and last mixture components are supposed to accommodate the under and overexpressed genes, respectively. The model is flexible enough to efficiently deal with the highly irregular spaced configuration of the data by assuming a hierarchical Markov dependence structure. The analysis of four cancer data sets (breast, lung, ovarian and uterus) is presented. Results indicate that the proposed model is selective in determining the regulation status, robust with respect to prior specifications and provides tools for a global or local search of under and overexpressed chromosome regions.
\vspace{3pt}
{\flushleft keywords: gene expression, mixture model, cancer, Gibbs sampling.}
\end{abstract}

\begin{figure}[b]
{ \flushleft \small \hspace{0.3cm}\emph{Address for correspondence:}
Vin\'{i}cius Mayrink, Departamento de Estat\'{i}stica, ICEx, UFMG. Av. Ant\^{o}nio Carlos, 6627, Belo Horizonte, MG, Brazil, 31270-901. E-mail: vdm@est.ufmg.br}
\end{figure}

%%%%%%%%%%%%%%%%%%%%%%
\section{Introduction}

Methods based on next-generation sequencing (NGS) to study the genome have flourished in the past years helping to understand changes in the transcriptome and leading to the discovery of new mutations and fusion genes. The identification of genes driving the cancer progression is, for example, the focus of extensive research using this technology \citep{Maher09,Berger10,Han11}. In this paper, we are particularly interested in the analysis of RNA-Seq (RNA sequencing) data quantifying the amount of RNA in a biological sample at a given moment in time \citep{Wang09,Oshlack10,Chu12,Conesa16}. The RNA-Seq data can be used in different types of investigations including, for example, modifications in gene expression over time \citep{Nueda14,Wiel13} and (more often) a differential expression analysis comparing distinct conditions, groups or tissue types \citep{Anders10,Bullard10,Robinson10,McCarthy12,Soneson13,Zhang15,Papas17}.

Prior to the advent of the NGS, the DNA microarrays \citep{Amaratunga14} were the major source of gene expression data explored in the literature. The first experiments to obtain RNA-Seq data were highly expensive compared to the microarray technology; this obstacle restricted the experimental design in many studies imposing analyses with few replications. The focus on the microarray technology is evident in the discussion presented in \cite{Horvath00} (back in 2000) about the perspectives of future directions in statistical genetics. As the NGS became affordable along the years, the number of studies based on RNA-Seq data increased rapidly leading to a scenario where it is now considered a powerful competitor to the microarrays. Studies confronting ``RNA-Seq versus microarrays'' can be easily found in the literature; see, for example, \cite{Marioni08} and \cite{Nookaew12}. The higher interest in RNA-Seq can also be explained by its advantages over the microarrays: lower background noise, the ability to detect low frequency and novel transcripts from genome regions not previously annotated, the requirement for less RNA samples to generate the data.

The work developed here is heavily based on the recent study reported in \cite{Mayrink17}, where the authors propose a Bayesian Markov mixture model to detect overexpressed regions on the chromosomes using microarray data. The model basically consists of a mixture of some gammas and one Gaussian distributions in which the latter is the mixture component with the highest mean and is suppose to accommodate the overexpressed regions. The authors use the known expected similarity among near locations as an important source of information by imposing a Markov structure to the expression measurements along the chromosome. Moreover, given the highly irregular distances between the observed locations, modelling and computational difficulties are overcome by a hierarchically structured Markov dependence as a function of the distances. Finally, the analysis is entirely focused on Affymetrix oligonucleotide arrays with the preprocessed (RMA) probe intensities being mapped to chromosome locations via the alignment tool BLAT.

Another interesting study dealing with the detection and classification of chromosome regions according to the status of the expressions from RNA-Seq can be found in \cite{Frazee14}. The main goal is the comparison between groups (e.g.: male/female or control/treatment) to evaluate the existence of expression differences. The authors fit a linear regression model (under the frequentist approach) to each aligned location in the genome, with the response being the number of sequences overlapping each location. The significance of coefficients (location specific and related to group indicators) is used to segment the genome into contiguous regions labeled as over, under or non-differentially expressed. The statistics to test the significance of such coefficients are determined conditionally on an underlying first-order Markov process with three states.

The approaches from \cite{Mayrink17} and \cite{Frazee14} differ in some critical aspects. In terms of modelling, \cite{Frazee14} does not consider the distance between locations as a source of information. In addition, the target application in \cite{Frazee14} is to compare groups and identify genomic regions exhibiting differences among those groups. The methodology proposed in \cite{Mayrink17}, on the other hand, aims at detecting high expression regions within a single group (e.g.: a tumor type) without having a reference group to evaluate distinctions. The overexpressed status in \cite{Mayrink17} is meant to call the attention for specific parts of the genome where the genes may be up-regulated and playing leading roles for the progression of the disease under investigation.

The main contribution of the present paper is to extend (considering reasonable adaptations) the methodology proposed in \cite{Mayrink17} for applications involving RNA-Seq data. As it will be discussed later, unlike the microarray data from \cite{Mayrink17}, the preprocessed observations considered here present a symmetric empirical behaviour. For that reason, the original gamma mixture components are replaced by Gaussian ones. Moreover, the present work considers not only the detection of overexpressed chromosome regions but also the detection of underexpressed ones. Finally, the proposed methodology is applied to four datasets, each of them concerning a different type of cancer (breast, lung, ovarian and uterus).

The outline of this paper is as follows. Section 2 describes the RNA-Seq data to be analysed, including all the preprocessing steps to organise, rescale and remove part of the noise in the raw data. Section 3 presents the proposed mixture model while Section 4 describes the details of the Bayesian inference procedure, including the MCMC algorithm and the cluster detection strategy. Section 5 shows the analysis of the four cancer data sets to illustrate the methodology. Finally, Section 6 presents the main conclusions and final remarks.

\vspace{-10pt}

%%%%%%%%%%%%%%%%%%%%%%
\section{The data} \label{secdata}

The RNA-Seq data, explored in this study, was obtained from the Genomic Data Commons (GDC) website maintained by the U.S. National Cancer Institute (NCI). According to their own description, the GDC data portal is a robust data-driven platform providing an unified repository that enables data sharing across cancer genomic studies; see \texttt{https://portal.gdc.cancer.gov} \, for more information. The data collected from the GDC portal represent four types of cancer: breast invasive carcinoma, lung adenocarcinoma, ovarian serous cystadenocarcinoma and uterine corpus endometrial carcinoma. Hereafter, these types of tumors will be denoted by: Breast (45 samples), Lung (56 samples), Ovarian (72 samples) and Uterus (48 samples), respectively. All samples are related to female human subjects with the age at diagnosis ranging between 40--90 years old; the death occurred within the period of 2 years since the diagnosis. Given that the data refers only to female subjects, the chromosome $Y$ is not included in our analysis.

Each of the samples is composed by thousands of measurements related to nucleotide sequences (genes) in the genome. The main steps to produce the data are: ($i$) the RNA's are isolated from a tissue and converted to cDNA fragments, ($ii$) a high-throughput sequencer generates millions of reads (short nucleotide sequences) from the cDNA fragments, ($iii$) an alignment tool is used to map the reads to the human genome and ($iv$) the number of reads mapped to a gene is regarded as the raw expression value.

The read mapping uncertainty can be seen as an experimental challenge in the process to build this type of data. The reads are shorter than the transcripts from which they are originated, thus a single read may map to multiple genes, which is certainly a factor complicating the analysis. There are many approaches proposed in the literature to deal with this problem; some examples are found in \cite{Marioni08}, \cite{Faulkner08}, \cite{Mortazavi08}, \cite{Li10}, \cite{Garber11} and \cite{Fonseca12}. The mapping uncertainty issue is not the focus of the present paper; the data sets extracted from the GDC portal have already been aligned to the human reference genome. For each sample, the data set contains: the gene identification (according the \textit{ensembl} annotation, see \texttt{http://www.ensembl.org}) and the number of reads mapped to that gene. Using the software \texttt{R} \citep{softwareR} and the package \texttt{Genominator} \citep{Bullard10} from the Bioconductor project (\texttt{http://bioconductor.org}, see \cite{Gentleman04}), one can easily retrieve the chromosome name and the start/end position of the gene within the chromosome by comparing the gene annotations (reported in the data sets) with the current version of the human genome (labeled as ``\texttt{hsapiens\_gene\_ensembl}'' in the \texttt{R} package). In particular, the start/end position of a gene provides two important quantities for our study; their average is assumed as the location of the gene and their difference indicates the length of the gene. Very few inconsistencies were found in the data - basically locations with multiple expressions; these cases were deleted from the analysis without significant loss. At this point of the data description, each sample (for any cancer) comprises $57{,}528$ genes

The raw expression measurements in a RNA-Seq data set must undergo preprocessing steps for normalisation before any statistical analysis. Normalisation is essential in gene expression studies using either microarray \citep{Mayrink15,Irizarry03} or RNA-Seq data \citep{Robinson10,Bullard10,Hansen12,Dillies12}. In the RNA-Seq case, the procedure is applied to remove systematic variations related to between-sample differences (i.e.: library size or sequencing depth, see \cite{Mortazavi08}) and within-sample effects (gene lengths, see \cite{Oshlack09}). More specifically, different samples may have different library sizes; i.e., their total number of mapped reads may differ. The bigger the library size, the larger are the counts observed for the genes; this implies in higher biological variability across the samples. In addition, it is well known that the genes have different lengths and, as a consequence, more reads tend to be mapped to the longer ones, misleading their expression with respect to the others.

Using the \texttt{R} package \texttt{edgeR} \citep{Robinson10} as a support, the following normalisation steps have been applied to the raw RNA-Seq data in this paper.
\begin{enumerate}
 \item Filtering: If a gene has $0$ counts for the majority of the samples, then it is not clear whether that gene is indeed unimportant or if it could not be properly detected by the sequencer.
          In line with this comment, genes with non-zero read counts in less than $10$ samples (for each cancer) were removed from the analysis. \vspace{-5pt}
 \item Resetting the library size: given the removal in the previous step, the total read count per sample is updated.  \vspace{-5pt}
 \item Computing the normalising factor: a value is obtained for each sample to scale the updated library sizes. The method Trimmed Mean of M-values (TMM) in \cite{RobOsh10} is considered here.  \vspace{-5pt}
 \item Normalising between- and within-samples: the method RPKM (reads per kilobase per million, implemented in \texttt{edgeR}) is applied taking as input arguments the normalising factor (previous step) and the gene lengths.
          The preprocessed expressions are reported in the $\log_2$ scale.
\end{enumerate}

It is important to emphasise that the group of genes removed in the filtering step of this normalisation is not the same for all cancers. As a result, the list and number of genes differ between the normalised data sets; we have: $37{,}752$ genes (Breast), $39{,}397$ (Lung), $44{,}293$ (Ovarian) and $38{,}739$ (Uterus). The total number of genes being evaluated in at least one cancer is $45{,}543$ ($77.3 \%$ of them are found in all four data sets). Table \ref{tab1} shows percentages indicating the level of intersections between the list of genes of two data sets. Note that the lowest percentage is $83.6\%$.

\begin{figure}[!h]
\centering
$$
 \begin{array}{cc}
  \mbox{\small (a)} & \hspace{-0.9cm} \mbox{\small (b)} \\
  \includegraphics[scale=0.25]{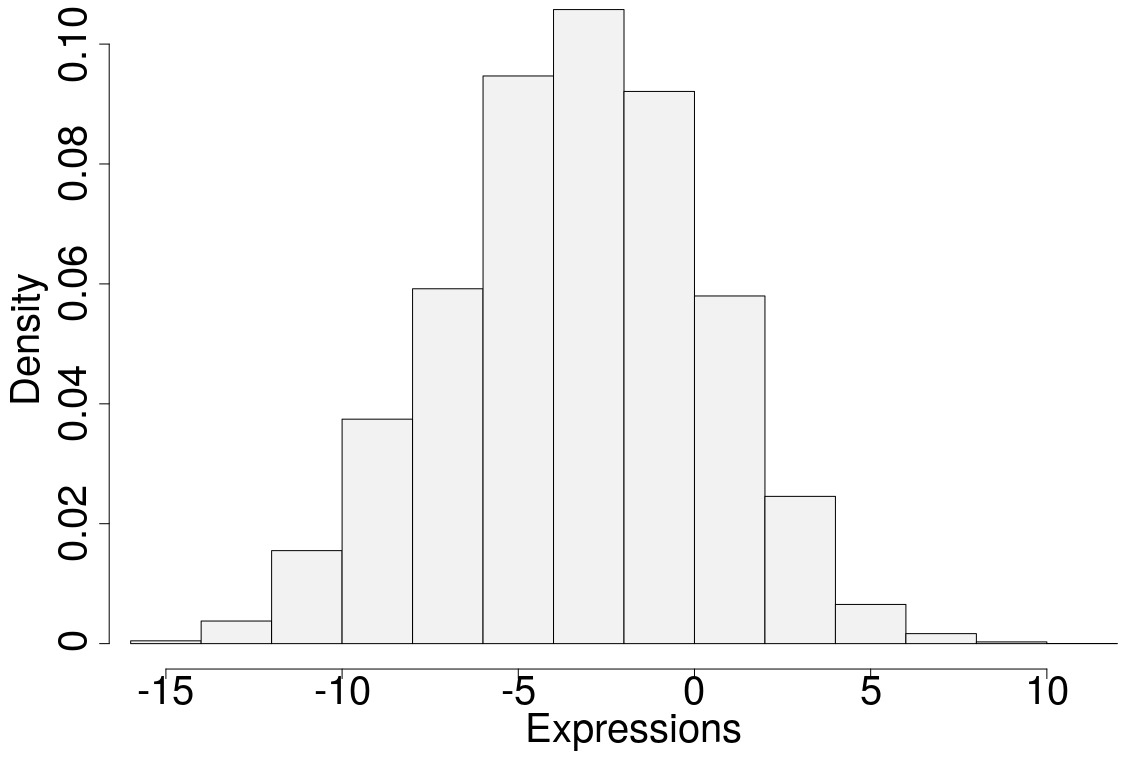} &
  \includegraphics[scale=0.22]{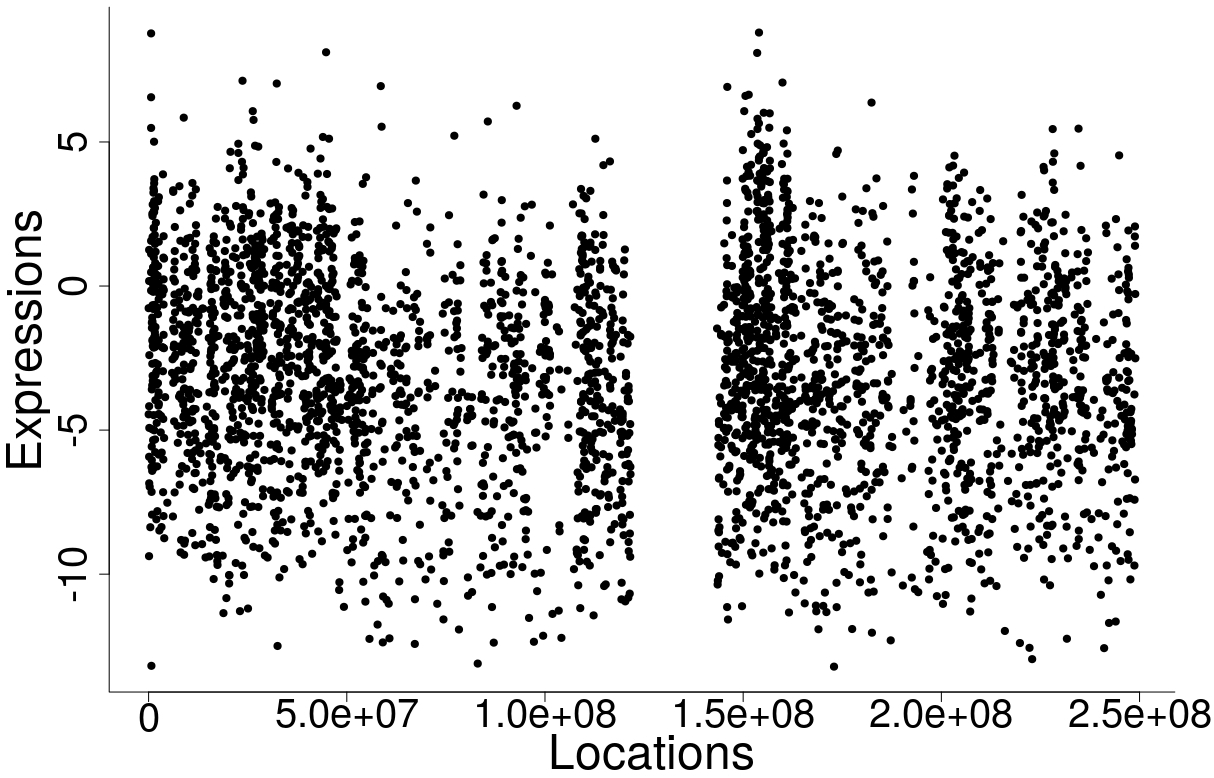} \\
 \end{array}
$$
\vspace{-10pt}
\caption{Graphs related to the data set ``Breast". Panel (a): Histogram displaying the distribution of the medians of the preprocessed $\log_2$ expressions (chromosomes $1-22$ and $X$). Panel (b): Points showing the positions of the medians along chromosome 1.}
\label{fig1}
\end{figure}

The results shown in Figure \ref{fig1} are based on a summarisation of the Breast data set using the median of the ``non-zero observations'' across the samples for each location in the genome (i.e., each gene). In order to simplify the discussion, the term ``non-zero observations'' is used here as a reference to the preprocessed values not associated with the zero counts in the raw data. The histogram in Panel (a) clearly has a symmetric shape contrasting with the skewed and multimodal distribution observed in \cite{Mayrink17} for microarray data. Panel (b) exhibits the spatial configuration of the data along the chromosome 1. Note that distances vary between data points and the graph exhibits the original scale of chromosome positions. The data associated with the other types of cancer have a very similar behaviour.

\begin{table}[h!]
\centering \small
\caption{Comparison involving pairs of data sets. Percentage of genes listed in both data sets of the pair relative to the number of genes listed in at least one of them.} \label{tab1}
\begin{tabular}{rrrrr}
\hline
& Breast & Lung & Ovarian & Uterus \\
\hline
Breast	& 100 &   90.6 &   83.6 &  87.8 \\
Lung		&        & 100    &   86.1 &  88.1 \\
Ovarian	&        &           & 100    &  85.3 \\
Uterus	&        &           &           & 100  \\
\hline
\end{tabular}
\end{table}

\begin{figure}[!h]
\centering
\includegraphics[scale=0.25]{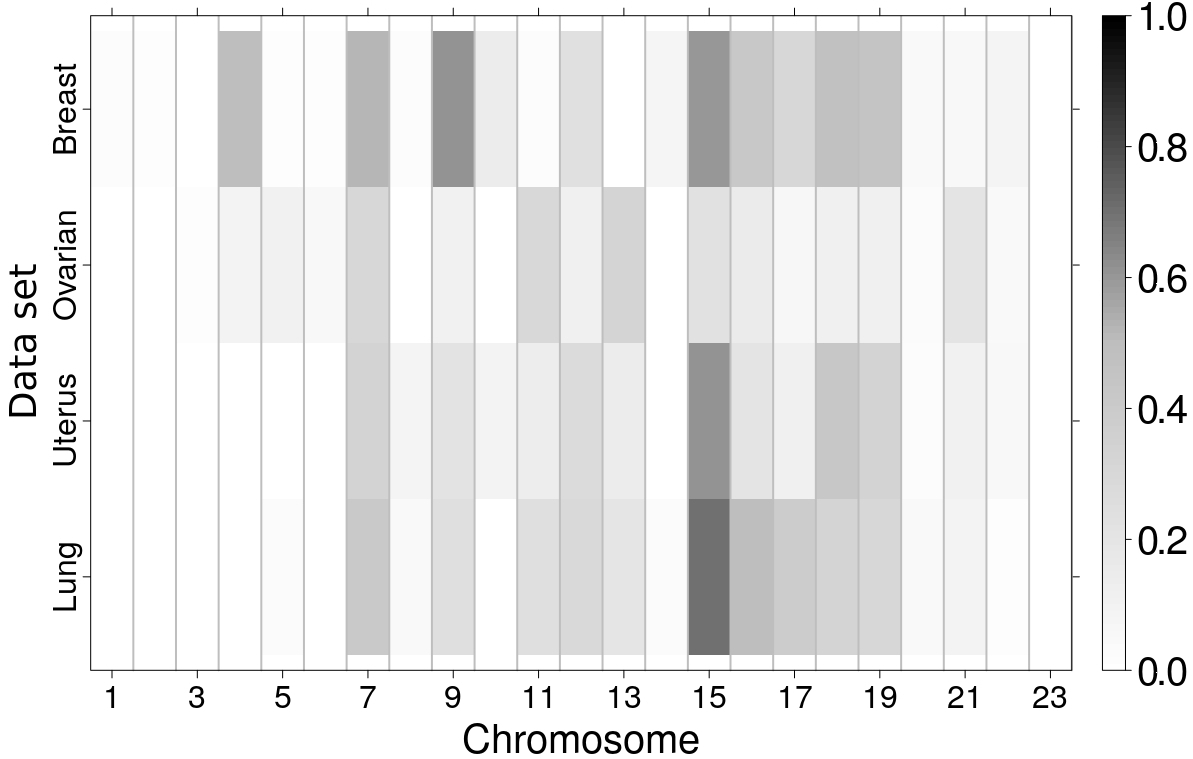}
\caption{Image displaying all p-values of the Moran's I permutation test applied to all data sets (summarised by the median); cancers in each row and chromosomes ($1-22$ and $X$) in each column. }
\label{fig2}
\end{figure}

In any cancer, the histograms for each preprocessed sample (i.e., without the median summarisation) indicate exactly the same symmetric shape shown in Figure \ref{fig1} with support also condensed between ($-15, 10$). This suggests that the overall behaviour observed in the median values is similar to the individual configuration in each sample. We choose to follow the same strategy developed in \cite{Mayrink17} to simplify the analysis by using the median (in our case, related to the non-zero counts across samples) to summarize the data. Recall that there are at least 10 non-zero counts for each gene, therefore, the calculation of the median does not rely on few values nor it is affected by the majority of zeros related to some genes. In addition, the distribution of the ``non-zero observations'' is in general unimodal and symmetric.

Figure \ref{fig2} shows a heat map graph displaying the p-values of the Moran's I permutation test for each chromosome and each cancer data (using the median summarisation). This is a well known test to verify the presence of spatial dependence in a data set; see \cite{Moran50} for details. The \texttt{R} package \texttt{spdep} \cite{Bivand15} is considered here to apply the test. As it can be seen, several p-values are small (in white) suggesting that the spatial association is significant in many cases. In addition, some large p-values (in grey) are also observed suggesting absence of spatial association. This combination of results definitely motivates a modelling strategy allowing for the presence or absence of spatial dependence and accounting for the distances between neighbours to explain this dependence.

Note that the magnitude of the original distances, indicated in Figure \ref{fig1} (b), is too large and this might determine computational difficulties to fit a model. This concern is also reported in \cite{Mayrink17} and the paper handles this problem by rescaling the distances to the interval ($0, 1$) using two steps: calculate the $\log$-distances and divide each of them by the maximum $\log$-distance. We will apply the same transformation in our study. As discussed in \cite{Mayrink17}, taking the $\log$ preserves high differences among the smallest values and practically even out the highest values; this is coherent with the model proposed in the next section, which assumes the distances as covariates to explain the Markov dependence.

\begin{figure}[!h]
\centering
\includegraphics[scale=0.35]{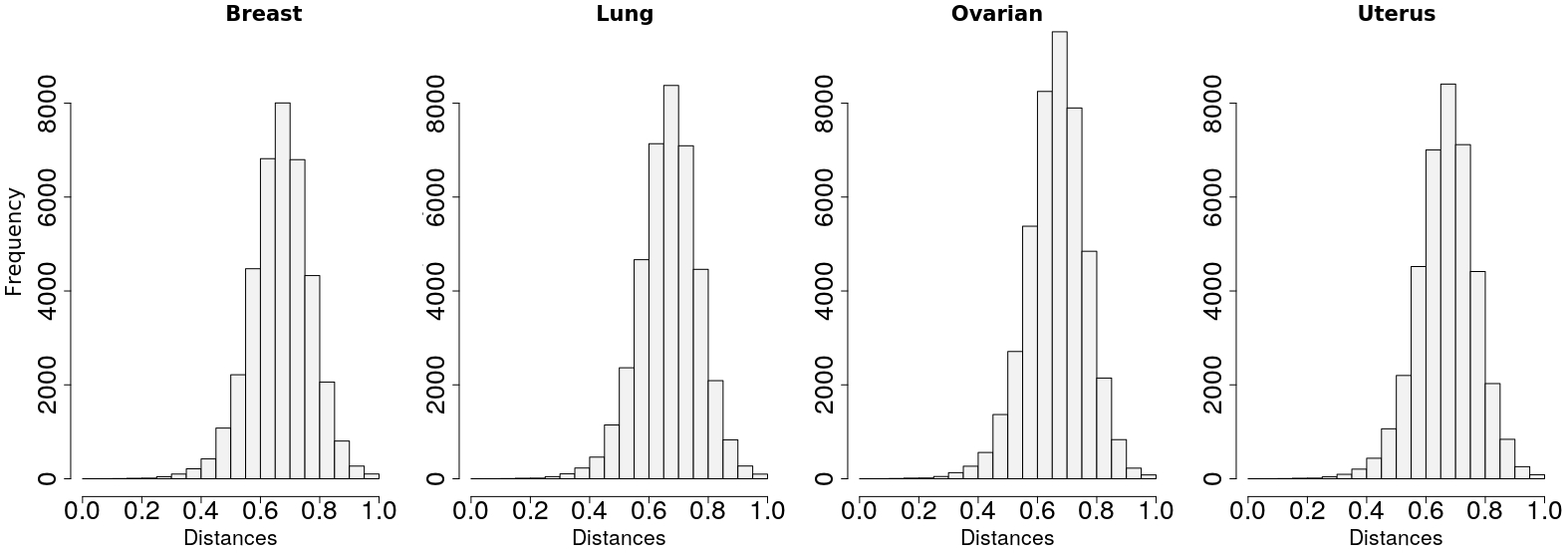}
\caption{Histograms of all rescaled distances between locations in all chromosomes.}
\label{fig3}
\end{figure}

Figure \ref{fig3} presents the histograms for all rescaled distances in the chromosomes. Note that most values are above $0.5$ and this result differs from the corresponding one in \cite{Mayrink17}, where the histogram has higher dispersion in the interval ($0, 1$). Two aspects can explain this difference. First, the number of genomic positions evaluated in \cite{Mayrink17} is more than twice the size of the largest data set in our study. Another reason is the fact that the probe sequences (not the bigger gene sequences) are the ones aligned to the genome in the microarray study; in our case, the positions represent genes. Probes are small fractions of a gene, therefore, the distance between two probes belonging to the same gene is potentially smaller than the distance between two genes. According to this discussion, it is reasonable to expect the result observed in Figure \ref{fig3}.

%%%%%%%%%%%%%%%%%%%%%%
\section{Bayesian hidden Markov mixture model} \label{secmodel}

For a cancer data set, let $n$ be the number of genes aligned to the genome (each gene has a unique location). Define the data as $X = \{ X_i, \; i = 1, \ldots, n \}$ where $X_{i}$ is the median of the preprocessed expressions in location $i$ that were not originally zero. Again, the summarisation through the median is justified by symmetry, unimodality and the number of replications available for this calculation.

The main purpose of this study is to detect underexpressed (down-regulated) and overexpressed (up-regulated) regions in the human genome for each selected cancer. Recall that \cite{Mayrink17} models only the overexpressed regions; the down-regulated cases are left aside in the analysis due to the strong right-skewed behaviour of the data. The underexpressed status of a gene is also a valuable piece of information that could help to explain key details about a disease. As shown in Figure \ref{fig1} (a), the preprocessed RNA-Seq data is fairly symmetric making both tails of the histogram equally important to be modelled - the left tail being connected to the down-regulation. It is not difficult to find in the literature studies evaluating both down and up-regulated status of a gene based on high-throughput sequencing data; a few examples are \cite{Han11}, \cite{Trapnell13} and \cite{Law14}.

The naive strategy of fixing two arbitrary thresholds and identifying the gene expressions (in the tails) below/above these values is discarded in our study for the same reasons discussed in \cite{Mayrink17}: ($i$) there is no explicit gap in the distribution of the expressions in each chromosome, ($ii$) there is no clear scale of the measured expression due to the normalisation steps and ($iii$) this attempt would ignore the dependence known to exist among near genes (e.g.: see DNA copy number alteration analysis in \cite{Pollack02} and \cite{Lucas10}).

Our model considers the dependence among near genes to find the chromosome regions of interest. An underexpressed (overexpressed) region is seen as a sequence of genes showing low (high) expressions in the chromosome. Such regions will be associated with the components having the smallest and largest mean in a mixture distribution; no fixed thresholds are specified here. The dependence structure is established by assuming that near genes are likely to be in the same mixture component. This structure also disfavors the detection of spikes, i.e., an isolated differentially expressed gene whose neighbours are not down or up-regulated as well. In order to achieve this goal, informative prior distributions are required for some parameters in the model.

Unlike \cite{Mayrink17} that considers a mixture of gamma distributions to efficiently account for the significant skewness of the microarray data explored in that work, this paper considers a mixture of Gaussian distributions due to the fairly symmetric behaviour of the RNA-Seq to be analysed. This change also simplifies the MCMC algorithm developed to perform inference.

Following the steps of \cite{Mayrink17}, we define a hierarchical Markov structure to model the dependence among the genes. On the first level, the Markov dependence is assumed to be the same for any consecutive pair of genes aligned to the genome, however, this dependence is present with a probability which is modelled as a function of the distances on the second level. The Markov dependence is essential to handle the fact that it is unlikely that a single location with an expression much lower/higher than its neighbours is indeed representing a gene from a differentially expressed region. The up/down spiked observations suggest a local atypical expression and they are expected to be captured by the left or right tail of the adjacent Gaussian component in the mixture. The cases where an isolated spike may belong to a potential differentially expressed region are those where their neighbours are not too close.

Let $K$ be the number of normal components in the mixture distribution. Note that $K \geq 3$ in our application, since the model must have Gaussian components to represent the under, over and non-differentially expressed categories. Define: the probability vector $q_0 = (q_{01},\ldots,q_{0K})'$ and the transition matrix $Q = \{q_{k_1k_2}\}$ of a $K$ states and discrete time Markov chain; $q_k$ is the $k$-th row of $Q$ and consider $k_1, k_2, k \, = \, 1, \ldots, K$. The $k$-th Gaussian component has mean $\mu_k$, variance $\sigma^2_k$ and its density will be represented by $f_k$. In addition, let $Z_i = (Z_{i,1},\ldots,Z_{i,K})'$ be a vector such that $Z_{i,k} = 1$ indicates that $X_i$ belongs to the $k$-th mixture component, $0$ otherwise. The following model is proposed:
\begin{eqnarray}
  (X_i|Z_{i,k}=1) & \sim & \mbox{N}(\mu_k, \sigma^2_k), \quad i = 1, \ldots, n,\;\mbox{all independent}; \label{meq1}\\
  (Z_1|q_0) & \sim & \mbox{Mult}(1, q_0);  \label{meq2} \\
  (Z_i|Z_{i-1,k}=1, \rho_i, q_0, Q) & \sim & \rho_i \; \mbox{Mult}(1, q_0) + (1-\rho_i) \; \mbox{Mult}(1, q_k), \quad i = 2, \ldots, n. \label{meq3}
\end{eqnarray}
The parameters indexing the mixture components are suppressed from the conditional notation above. The term ``Mult'' refers to the multinomial distribution. The distribution in (\ref{meq3}) implies that a Markov dependence between the expressions in the locations $i-1$ and $i$ exists with probability $\rho_i$.

As opposed to an ordinary mixture model, the $X_i$ variables are not marginally (w.r.t. $Z$) independent in (\ref{meq1})-(\ref{meq3}). The latent level (\ref{meq3}) defines the dependence through a Markov structure. For computational reasons, the Bernoulli random variables $W_i$, $i=2,\ldots,n$, are introduced in the model as indicators of ``presence of Markov dependence''. In this case, consider for $i = 2, \ldots, n$:
\begin{eqnarray}
  (Z_i|Z_{i-1,k}=1,W_i=0,q_0) & \sim & \mbox{Mult}(1,q_0); \label{meq4} \\
  (Z_i|Z_{i-1,k}=1,W_i=1,Q) & \sim & \mbox{Mult}(1,q_k); \label{meq5} \\
  (W_i|\rho_i) & \sim & \mbox{Ber}(\rho_i). \label{meq6}
\end{eqnarray}
Assume that $W_1 = 0$ almost surely and $(Z_1|Z_0,W_1=0,q_0) := (Z_1|q_0)$.

The full specification of the Bayesian model is completed with the prior distributions:
\begin{eqnarray}
  q_0 & \sim & \mbox{Dir}(r_0); \nonumber \\
  q_k & \sim & \mbox{Dir}(r_k); \; \mbox{all independent}; \nonumber \\
  (\mu_k,\sigma^2_k) & \sim & \mbox{NIG}(m_k,v_k,s_{1k},s_{2k}), \quad \mu_1 < \mu_2 < \ldots < \mu_K; \label{PriorDist} \\
  (\rho_i|\beta) & = & \Phi(\beta_0+\beta_1d_i); \nonumber\\
  \beta = (\beta_0,\beta_1)' &\sim & \mbox{N}_2(\mu_0,\Sigma_0). \nonumber
\end{eqnarray}
In the specifications above, $\Phi(.)$ is the c.d.f. of the standard normal distribution and $d_i$ is the (transformed) distance between locations $i-1$ and $i$. The terms Dir, NIG and $\mbox{N}_2$ refer to the Dirichlet, normal-inverse-gamma and bivariate normal distributions, respectively. Additional notation: $Z = \{ Z_i, \; i = 1, \ldots, n \}$, $W = \{W_i, \; i = 1, \ldots, n \}$,  $\mu = \{ \mu_k, \; k=1, \ldots, K \}$ and $\sigma^2 = \{\sigma^2_k, \; k = 1, \ldots, K \}$.

Note that the distance $d_i$ is used to stochastically explain the Markov dependence through the probit regression on $W_i$. Vector $q_0$ could be seen as the stationary distribution of the Markov chain with transition matrix $Q$; however, this restriction is not imposed in our model, since it would complicate the inference methodology.

The lower ($k = 1$) and upper ($k = K$) Gaussian components of the mixture are expected to accommodate the expressions of the genes more likely to form the under and overexpressed regions of interest, respectively. Since the percentages of genes in those regions are expected to be small, one can elicit informative prior distributions to insert this idea; for example, priors suggesting small values for $\{q_{0,1},q_{1,1},\ldots,q_{K,1}\}$ and $\{q_{0,K},q_{1,K},\ldots,q_{K,K}\}$.

%%%%%%%%%%%%%%%%%%%%%%
\section{Bayesian inference}

In this section, we present an MCMC algorithm to sample from the high dimensional and complex joint posterior distribution of all unknown quantities defined in the proposed model. The implementation was developed using the \texttt{R} programming language \citep{softwareR}. Unlike the MCMC algorithm from \cite{Mayrink17}, no Metropolis-Hastings steps are required in our Gibbs sampler due to the use of only Gaussian mixture components.

%%%%%%%%%%%%%%%%%%%%%%
\subsection{The MCMC algorithm} \label{ssecmcmc}

The algorithm is a Gibbs sampling with blocking and sampling schemes chosen in a way to favor fast convergence of the chain. In this direction, consider a set of independent auxiliary variables $V := \{ V_i, \; i = 2, \ldots, n \}$ to allow for direct sampling from the full conditional distribution of $\beta$ (see \cite{Albert92}). Let $\vec{d}_i = (1, d_i)'$ and define:
\begin{equation} \label{auxvrb}
\ds (V_i|\beta) \sim N(\beta' \vec{d}_i, 1) \quad \mbox{and} \quad
    W_i \, = \, \left\{ \begin{array}{ll}
                 1, & \mbox{if} \; V_i>0 \\
                 0, & \mbox{if} \; V_i \leq 0.
            \end{array} \right. \nonumber
\end{equation}
The original model for $X$ is preserved here; this can be verified by simply integrating $V$ out.

Let $\psi := \{ \mu, \sigma^2 \}$ and consider the following blocking scheme for the Gibbs sampler: \, $(Z,W)$, \, $V$, \, $(q_0,Q,\beta,\psi)$. In order to define an irreducible Markov chain, we consider a collapsed Gibbs sampling (see \cite{Liu94}) with $V$ being integrated out from the full conditional distribution of $(Z,W)$.

All full conditional densities are proportional to the joint density of $X$ and the unknown quantities in the model - this density is given in Appendix A (\ref{jdens}). The next items describe each step of the algorithm.

\newpage

\noindent \textit{Sampling $V$:} \vspace{3pt}

The $V_i$'s are all independent with distribution $\ds \mbox{N}(\beta'\vec{d}_i, 1)$, truncated to be positive if $W_i = 1$ and non-positive if $W_i = 0$.

\vspace{5pt}

\noindent \textit{Sampling $(q_0, Q, \beta, \psi)$:} \vspace{3pt}

The four elements in this block - $q_0$, $Q$, $\beta$ and $\psi$, are conditionally independent; therefore, each of them can be sampled individually from the corresponding marginal full conditional distribution. The marginals of the first three components are given in Appendix A (\ref{fc1})-(\ref{fc3}). The marginal full conditional of $\psi$ is a truncation of the tractable distribution shown in Appendix A (\ref{fc4}); the truncation region is defined by the restriction $\{\mu_1 < \mu_2 < \ldots <\mu_K\}$. Using the rejection sampling, one can sample exactly from this truncated distribution by proposing from its non-truncated version and accepting if the restriction is preserved. Since $K$ is chosen to be small, the algorithm is computationally efficient.

\vspace{5pt}

\noindent \textit{Sampling $(Z,W)$:} \vspace{3pt}

This step of the algorithm was developed in \cite{Mayrink17} and consists of a backward-filtering-forward-sampling (BFFS) scheme. Note that integrating out $V$ provides the following full conditional kernel:
\begin{eqnarray}\label{fcZW}
\ds \pi(Z,W|\cdot) & \propto & \prod_{i=1}^n\left[\prod_{k=1}^{K}\left[ (f_k(X_i|\psi)q_{0k})^{Z_{ik}}\right]^{1-W_i} \left[(f_k(X_i|\psi)q_{k_{(i-1)}k})^{Z_{ik}}\right]^{W_i}\right] \nonumber \\
& & \times \; \left( \Phi_i^+ \right)^{W_i}\left( \Phi_i^- \right)^{1-W_i}\prod_{k=1}^{K}q_{0k}^{Z_{0k}},
\end{eqnarray}
with $\Phi_i^+ = \Phi(\beta'\vec{d}_i)$ and $\Phi_i^- = \Phi(-\beta'\vec{d}_i)$.

Direct sampling from (\ref{fcZW}) is possible if we assume the factorisation:
\begin{equation}\label{fcZ2}
\ds \pi(Z,W|\cdot)\propto\pi(Z_1|\cdot)\prod_{i=2}^n\pi(Z_i|W_i,Z_{i-1},\cdot)\pi(W_i|Z_{i-1},\cdot). \nonumber
\end{equation}
This implies that $(Z,W)$ must be sampled in the forward direction, i.e., according to the order: $Z_1$, $W_2$, $Z_2$, $\ldots$, $W_n$, $Z_n$. The marginal distributions are multinomial (Bernoulli for $W_i$) and their parameter values are obtained recursively (backwards from $n$ to $1$). Further details about the BFFS are given in Appendix A.

\vspace{-5pt}

%%%%%%%%%%%%%%%%%%%%%%
\subsection{Cluster detection} \label{ssecclus}

As explained in Section \ref{secmodel}, the genes forming a cluster of down and up-regulation are supposed to be incorporated by the lower and upper Gaussian components of the mixture, respectively. Hence, the cluster detection ought to be done based on the posterior probability that one or more genes are accommodated by the (lower or upper) Gaussian component.

Given a sample of size $M$ from the posterior distribution of $Z$ and a sequence $\{i_1,i_2,\ldots,i_S; \, S = 1, 2, \ldots \}$ of expressions, the posterior probability of this sequence being a down-regulated or an up-regulated cluster of genes are, respectively:
\begin{equation} \label{clustprob}
\ds \frac{1}{M}\sum_{m=1}^M \prod_{s=1}^S \mathds{1}(Z_{i_s,1}^{(m)}=1) \hspace{1cm} \mbox{and} \hspace{1cm} \frac{1}{M}\sum_{m=1}^M \prod_{s=1}^S \mathds{1}(Z_{i_s,K}^{(m)}=1);
\end{equation}
where $\mathds{1}(.)$ is the indicator function.

In another possible criterion for cluster detection, one may find all genes with probability greater than a threshold (for example $0.5$) of belonging to a (lower or upper) Gaussian component; then, search which of them form contiguous sequences of a minimum size (say 4 or 5).

%%%%%%%%%%%%%%%%%%%%%%
\section{Results for the cancer data sets} \label{secresult}

The analysis reported in this section is related to four RNA-Seq data sets representing different types of cancers; see details in Section \ref{secdata}. Since the scale of the preprocessed expressions is the same for all chromosomes, the analysis is developed by organising the chromosomes $1-22$ and $X$ side by side (for each data set) to form the genomic sequence. Note that there exists no distance to be measured between the last and first locations of consecutive chromosomes, this means that we have no reason to assume a dependence between the gene expressions in these extremities. This aspect is incorporated in the model by fixing $W_i = 0$ for the first gene location in each chromosome. This assumption is computationally convenient since the BFFS procedure can be executed separately (therefore in parallel) for each chromosome due to conditional independence.

Choosing the number of Gaussian components in the mixture is certainly a critical step in our analysis. As discussed in Section \ref{secmodel}, $K$ is at least $3$; however, assuming $K$ small leads to a model fit where the lower/upper Gaussian components are too wide (with high variance) and include too many expressions. In order to explore other reasonable choices, the models with $K = 3$, $4$, $5$, $6$, $7$, $8$ and $9$ are fitted to the Breast and Ovarian data sets. Results are presented in Appendix B and they indicate that $K = 8$ is a suitable global choice showing the same behaviour between data sets and with respect to $K = 9$.

In terms of prior specification, we set {\small $r_0 = (32.26, 322.58, 322.58, 322.58, 322.58, 322.58, 322.58, 32.26)'$}, which indicates that an expression at a location without the Markov dependence most likely belongs to an internal (not the lower or upper) Gaussian component in the mixture; note that $r_{01} = r_{08} = r_{0k}/10$ for any $k \in \{ 2, \ldots, 7\}$. The sum of these weights is $2{,}000$ and smaller concentration parameters would determine weakly informative priors w.r.t. the data involving thousands of genes. The following matrix shows the prior specification for $q_k$.

{\footnotesize
$$
r = \left(
\begin{array}{cccccccc}
 1212.12 &   606.06 & 121.21 &   12.12 &   12.12 &   12.12 &     12.12 &     12.12 \\
     12.12 & 1212.12 & 606.06 & 121.21 &   12.12 &   12.12 &     12.12 &     12.12 \\
       9.35 &   467.29 & 934.58 & 467.29 &   93.46 &     9.35 &       9.35 &       9.35 \\
       8.97 &     89.69 & 448.43 & 896.86 & 448.43 &   89.69 &       8.97 &       8.97 \\
       8.97 &       8.97 &   89.69 & 448.43 & 896.86 & 448.43 &     89.69 &       8.97 \\
       9.35 &       9.35 &     9.35 &   93.46 & 467.29 & 934.58 &   467.29 &       9.35 \\
     12.12 &     12.12 &   12.12 &   12.12 & 121.21 & 606.06 & 1212.12 &     12.12 \\
     12.12 &     12.12 &   12.12 &   12.12 &   12.12 & 121.21 &   606.06 & 1212.12 \\
\end{array}
\right).$$}

The $k$-th row of $r$ contains the parameter vector for the Dirichlet prior of $q_k$. Note that the sum of any row is $2{,}000$ and the largest weights are defined in the main diagonal (the $r_{kk}$ entries). This configuration favors the allocation of the $i$-th observation in the same Gaussian component of the neighbour expression from location $i-1$. In addition, the weight in each row decreases when moving away from the main diagonal; in any row, the second and third largest weights are given by $r_{kk}/2$ and $r_{kk}/10$, respectively. This specification discourages the model to assign two consecutive observations in distant components. Small weights are also specified for the lower/upper Gaussian components in columns $1$ and $8$ (except for the main diagonal elements); we have $r_{k1} = r_{k8} = r_{kk}/100$, for $k = 2, \ldots, 7$. This choice indicates that the lower/upper components are accessible only to those genes with great evidence of differential expression and having neighbours with the same characteristic. We emphasise that $r_0$ and $r$ can be easily adapted to other values of $K$, such as those explored in Appendix B. All adaptations have the same discussed features: arrangement of small/large values, magnitude of weights (largest one divided by $2$, $10$ or $100$) and the sum of weights being $2{,}000$.

The coefficients ($\beta_0, \beta_1$) control the impact of the distance $d_i$ over the probability of Markov dependence $\rho_i$. The analysis in \cite{Mayrink17} estimates those parameters to be around ($5.05$,$-9.55$), based on a large number of expressions aligned to the human genome. The genomic sequence, in that case, is denser than those from our RNA-Seq data. Given this scenario and the fact that the distance scale is the same as in the present paper, it seems reasonable to assume that those estimates represent well the relationship between $d_i$ and $\rho_i$. Therefore, they ought to be used to elicit an informative prior. In line with that, we set $(\beta_0,\beta_1)' \sim \mbox{N}_2[ (5.05, -9.55)', 0.0001 \,\bm{I_2}]$, which indicates: $\lim_{d_i \rightarrow 0} \rho_i \approx 1$, $\lim_{d_i \rightarrow 1} \rho_i \approx 0$ and $\rho_i = 0.5$ for $d_i \approx 0.53$.

In the normal-inverse-gamma prior for ($\mu_k, \sigma^2_k$), we choose the mean $m_k$ based on the scale of the observations and the role expected for the Gaussian components in the model. In this case, let $v_k = 10$ for all $k$ and $m_k = -10$, $-7.14$, $-4.29$, $-1.43$, $1.43$, $4.29$, $7.14$ and $10$ for $k = 1, \ldots, 8$, respectively. We also set $s_{1k} = 2.1$ and $s_{2k} = 1.1$ for all $k$, meaning that $E(\sigma^2_k) = 1$ and $\mbox{Var}(\sigma^2_k) = 10$. A sensitivity analysis was performed by increasing the prior variance of ($\mu_k, \sigma^2_k$) and the very same results were obtained -- see Appendix B.

For each data set, the MCMC chain runs for $10{,}000$ iterations with a burn-in of $2{,}000$. In terms of initial values, we set $q_0^{(0)} = 0.125 \, \bm{1_{(8 \times 1)}}$ and $Q^{(0)} = 0.125 \, \bm{1_{(8 \times 8)}}$; where $\bm{1_{(l_1 \times l_2)}}$ is a $l_1 \times l_2$ matrix of ones. In addition, $\beta_0^{(0)} = 5.05$ and $\beta_1^{(0)} = -9.55$. The starting values of $\mu_k$ and $\sigma^2_k$ are set based on descriptive statistics exploring the data. The support of the histogram in Figure \ref{fig1} (a) is broken into 8 contiguous intervals and their means and variances are used to initialise the MCMC; Table \ref{tab2} shows the starting values for each data set. Convergence is rapidly attained -- Appendix C shows some diagnostics.

\begin{table}[h!]
\centering \small
\caption{Initial values in the MCMC for $\mu_k$ and $\sigma^2_k$ ($K = 8$ mixture components).}  \label{tab2}
\begin{tabular}{rrrrrrrrrr}
\hline
                & $k$ & 1 & 2 & 3 & 4 & 5 & 6 & 7 & 8 \\
\hline
Breast	& $\mu_k$ &-13.85 & -10.64 & -7.43 & -4.23 & -1.02 &  2.19 & 5.40 & 8.60 \\
		& $\sigma^2_k$ &   0.51 &    0.69 &   0.84 &  0.80 &  0.81 &  0.74 & 0.69 & 0.49 \\
\hline
Lung 	& $\mu_k$ & -13.40 & -10.11 & -6.83 & -3.54 & -0.26 & 3.03 & 6.31 & 9.59 \\
                & $\sigma^2_k$ &   0.46  &    0.76 &  0.89 &  0.85 &  0.86  & 0.74 & 0.73 & 0.77 \\
\hline
Ovarian	& $\mu_k$ & -14.12 & -10.82 & -7.52 &  -4.22 & -0.93 & 2.37 & 5.67 & 8.97 \\
                & $\sigma^2_k$ & 0.41 & 0.74 & 0.88 & 0.85 & 0.83 & 0.78 & 0.71 & 0.57 \\
\hline
Uterus	& $\mu_k$ & -12.88 & -9.77 & -6.65 & -3.54 & -0.42 &  2.69 &  5.81 &  8.92 \\
                & $\sigma^2_k$ & 0.42 & 0.67 & 0.80 & 0.76 & 0.76 & 0.70 & 0.68 & 0.50 \\
\hline
\end{tabular}
\end{table}

\begin{table}[h!]
\centering \small
\caption{Posterior mean and standard deviation (in parentheses) for the parameters in the mixture.} \label{tab3}
\begin{tabular}{rrrrr}
  \hline
& Breast & Lung & Ovarian & Uterus \vspace{3pt} \\
\hline
$\mu_1$ & -11.168 (0.162) & -11.019 (0.141) & -11.157 (0.130) & -10.426 (0.138) \\
$\sigma^2_1$ & 2.125 (0.200) & 1.789 (0.165) & 1.914 (0.158) & 1.766 (0.156) \vspace{3pt} \\
$\mu_2$ & -7.866 (0.100) & -7.707 (0.096) & -7.873 (0.088) & -7.244 (0.100) \\
$\sigma^2_2$ & 2.876 (0.179) & 2.965 (0.193) & 2.858 (0.186) & 2.638 (0.168) \vspace{3pt} \\
$\mu_3$ & -4.765 (0.080) & -4.569 (0.098) & -5.130 (0.089) & -3.974 (0.080) \\
$\sigma^2_3$ & 1.376 (0.112) & 1.479 (0.123) & 1.451 (0.150) & 1.271 (0.135) \vspace{3pt} \\
$\mu_4$ & -3.534 (0.314) & -3.375 (0.271) & -3.664 (0.075) & -3.652 (0.193) \\
$\sigma^2_4$ & 5.408 (0.776) & 4.028 (0.677) & 1.038 (0.129) & 4.514 (0.773) \vspace{3pt} \\
$\mu_5$ & -2.441 (0.150) & -2.293 (0.141) & -1.853 (0.049) & -2.152 (0.122) \\
$\sigma^2_5$ & 2.146 (0.331) & 1.858 (0.237) & 0.952 (0.083) & 1.544 (0.223) \vspace{3pt} \\
$\mu_6$ & -0.167 (0.180) & -0.049 (0.111) & -0.058 (0.082) & -0.064 (0.191) \\
$\sigma^2_6$ & 2.948 (0.336) & 2.467 (0.195) & 2.340 (0.164) & 2.557 (0.260) \vspace{3pt} \\
$\mu_7$ & 0.124 (0.177) & 0.256 (0.163) & 0.077 (0.105) & 0.670 (0.259) \\
$\sigma^2_7$ & 9.064 (0.544) & 9.364 (0.515) & 10.543 (0.425) & 8.817 (0.758) \vspace{3pt} \\
$\mu_8$ & 2.666 (0.312) & 2.843 (0.179) & 2.802 (0.197) & 2.920 (0.244) \\
$\sigma^2_8$ & 1.436 (0.350) & 1.282 (0.272) & 1.506 (0.339) & 1.230 (0.262) \\
\hline
\end{tabular}
\end{table}

\begin{table}[h!]
\centering \small
\caption{Posterior weights of the mixture components.}  \label{tab4}
\begin{tabular}{rrrrrrrrr}
\hline
$k$      & 1 & 2 & 3 & 4 & 5 & 6 & 7 & 8 \\
\hline
Breast & 0.026 & 0.180 & 0.153 & 0.160 & 0.163 & 0.162 & 0.143 & 0.014 \\
Lung & 0.025 & 0.186 & 0.155 & 0.159 & 0.167 & 0.155 & 0.138 & 0.014 \\
Ovarian & 0.028 & 0.180 & 0.161 & 0.164 & 0.162 & 0.157 & 0.135 & 0.014 \\
Uterus & 0.029 & 0.185 & 0.151 & 0.164 & 0.168 & 0.157 & 0.131 & 0.014 \\
\hline
\end{tabular}
\end{table}

\begin{figure}[!h]
\centering
   \includegraphics[scale=0.26]{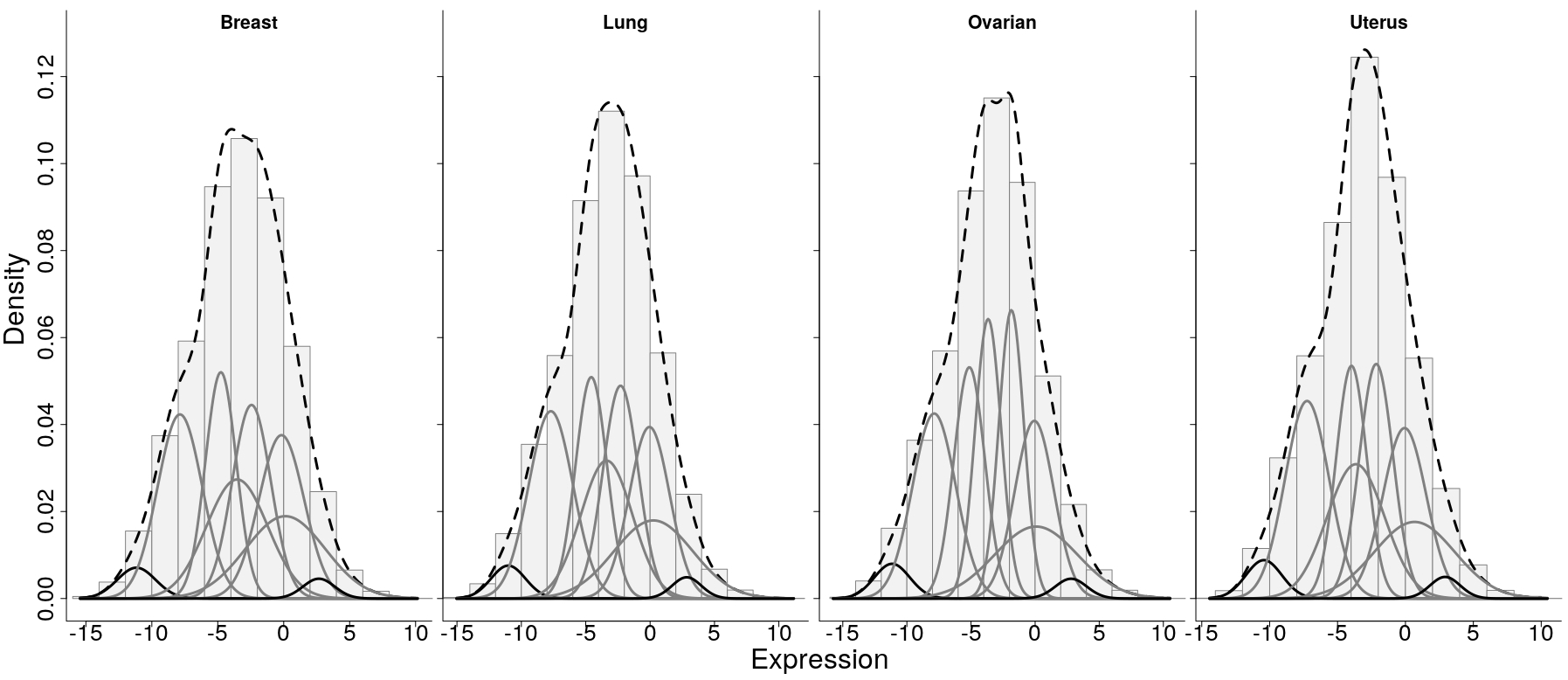}
\caption{Histogram of all expressions overlaid by the estimated mixture density (dashed curve) and its components (grey and black curves).}
\label{fig4}
\end{figure}

Table \ref{tab3} presents the posterior estimates of ($\mu_k, \sigma^2_k$) for all mixture components. Note that the values of the mean and standard deviation are similar across the different data sets, implying similar mixture densities. Figure \ref{fig4} confirms this visual interpretation by showing the data histograms overlaid by the estimated mixture density and their components. Note that the range of the lower/upper Gaussians are not too wide; therefore, they accommodate well the expressions in the left/right tails of the histograms. According to this result, the mixture with $K = 8$ components seems a good representation of the data. Table \ref{tab4} shows the posterior weight of each $k$ in each cancer; this value is the posterior mean of $\sum_{i = 1}^n Z_{ik}/n$.

\begin{figure}[!h]
\centering
   \includegraphics[scale=0.32]{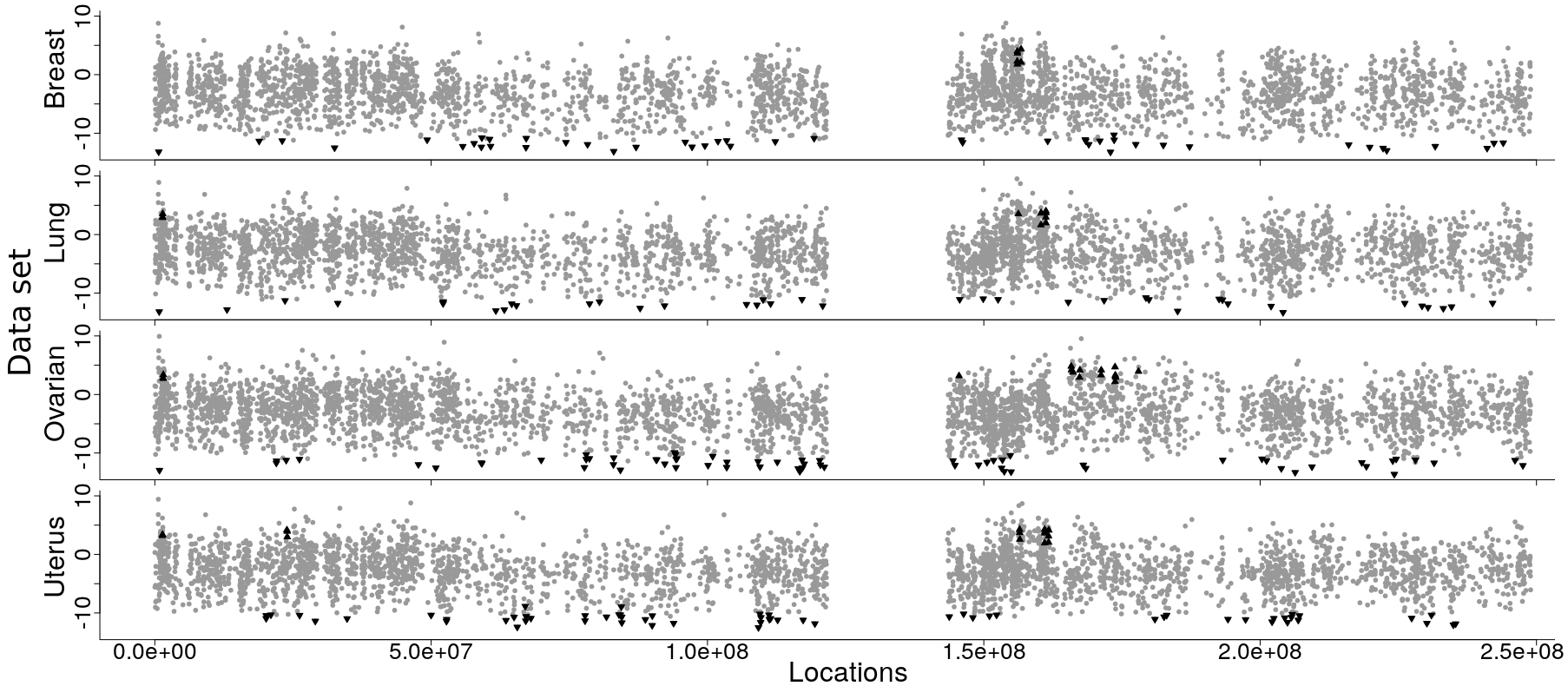}
\caption{Expression values along chromosome 1. The triangles indicate observations where the Gaussian component has posterior probability above $0.5$.}
\label{fig5}
\end{figure}

Figure \ref{fig5} shows the behaviour of the expression values along chromosome 1 for all data sets. As it can be seen, the black triangles, representing observations with high posterior probability ($> 0.5$) of belonging to the lower/upper Gaussian components, are concentrated on the bottom and top of the graphs. For chromosome 1, the lower Gaussian component clearly accommodates more expressions than the upper case. Note that internal Gaussian observations (grey dots) can also be identified in the bottom/top level. These expressions are allocated in adjacent Gaussian components, which is a consequence of the model structure accounting for distances between genes to establish the strength of the Markov dependence.

Some interesting similarities among the data sets can be observed in Figure \ref{fig5}; there are regions in chromosome 1 with the same under or overexpressed status in two data sets. In the analysis of the whole genomic sequence ($45{,}543$ locations in $23$ chromosomes), the number of genes allocated with high probability in the lower and upper Gaussian components are, respectively: $722$ and $31$ (Breast), $717$ and $54$ (Lung), $933$ and $74$ (Ovarian) and $877$ and $46$ (Uterus). Table \ref{tab5} cross-compares data sets and indicates the proportion of locations identified in the same Gaussian component (lower or upper) in both cancers, relative to the number of locations identified in that Gaussian component in at least one of the two cancers; this analysis is restricted to those genes represented in all four data sets. Note that the largest percentages are obtained in the Lung/Uterus comparison for both down and up-regulation status. Most percentages are $0$ in the upper Gaussian case, since the number of overexpressed genes identified by the model is small. On the other hand, the smallest percentage found in the lower Gaussian case is $1.40\%$ representing the Breast/Ovarian comparison.

\begin{figure}[!h]
$$
  \begin{array}{cc}
   \mbox{\small (a)} & \mbox{\small (b)} \\
   \includegraphics[scale=0.16]{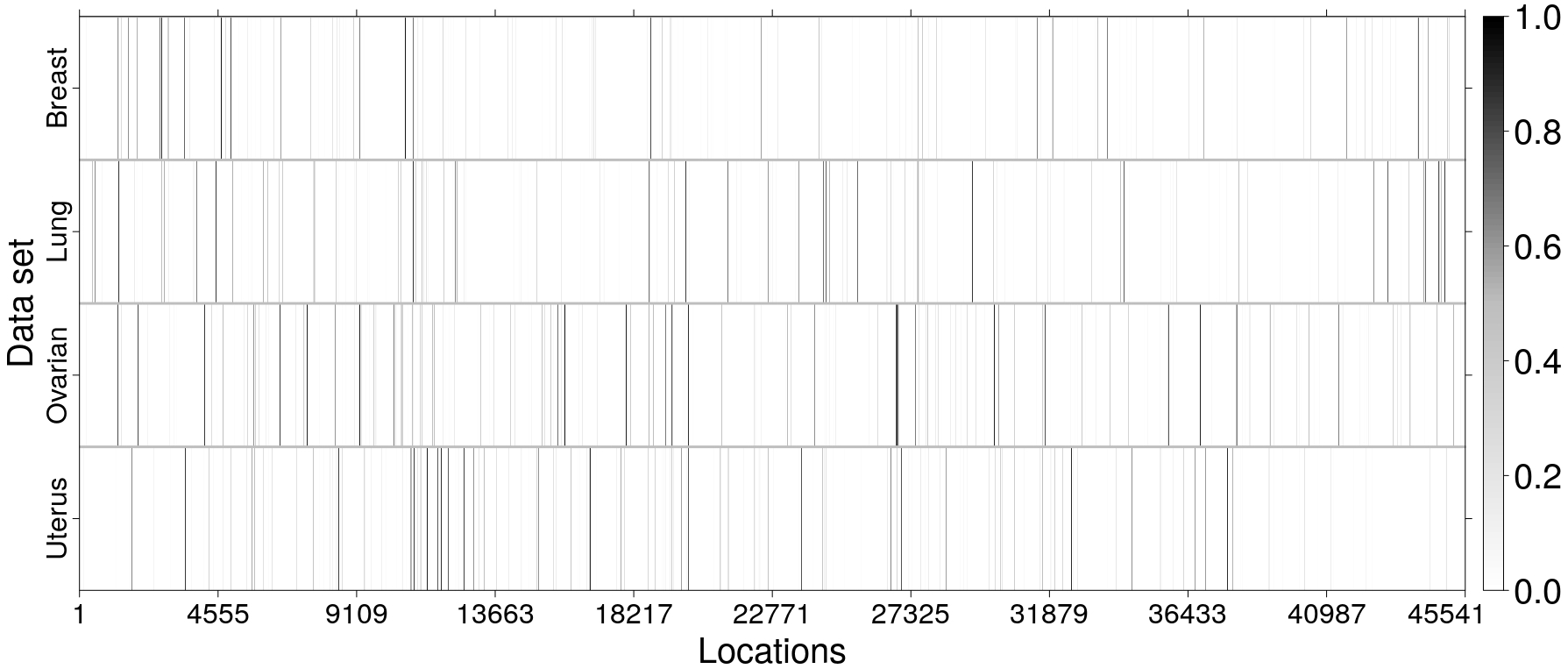} & \includegraphics[scale=0.16]{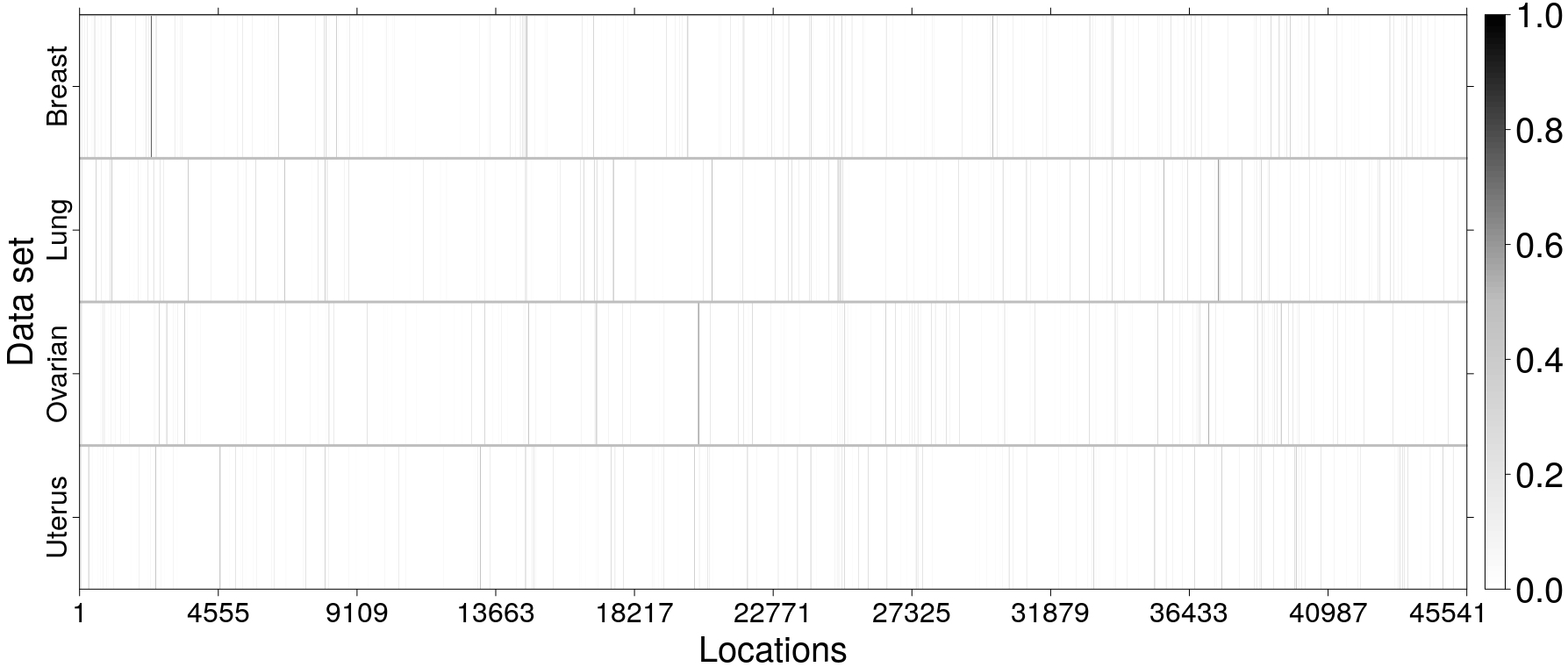}\\
   \mbox{\small (c)} & \mbox{\small (d)} \\
   \includegraphics[scale=0.16]{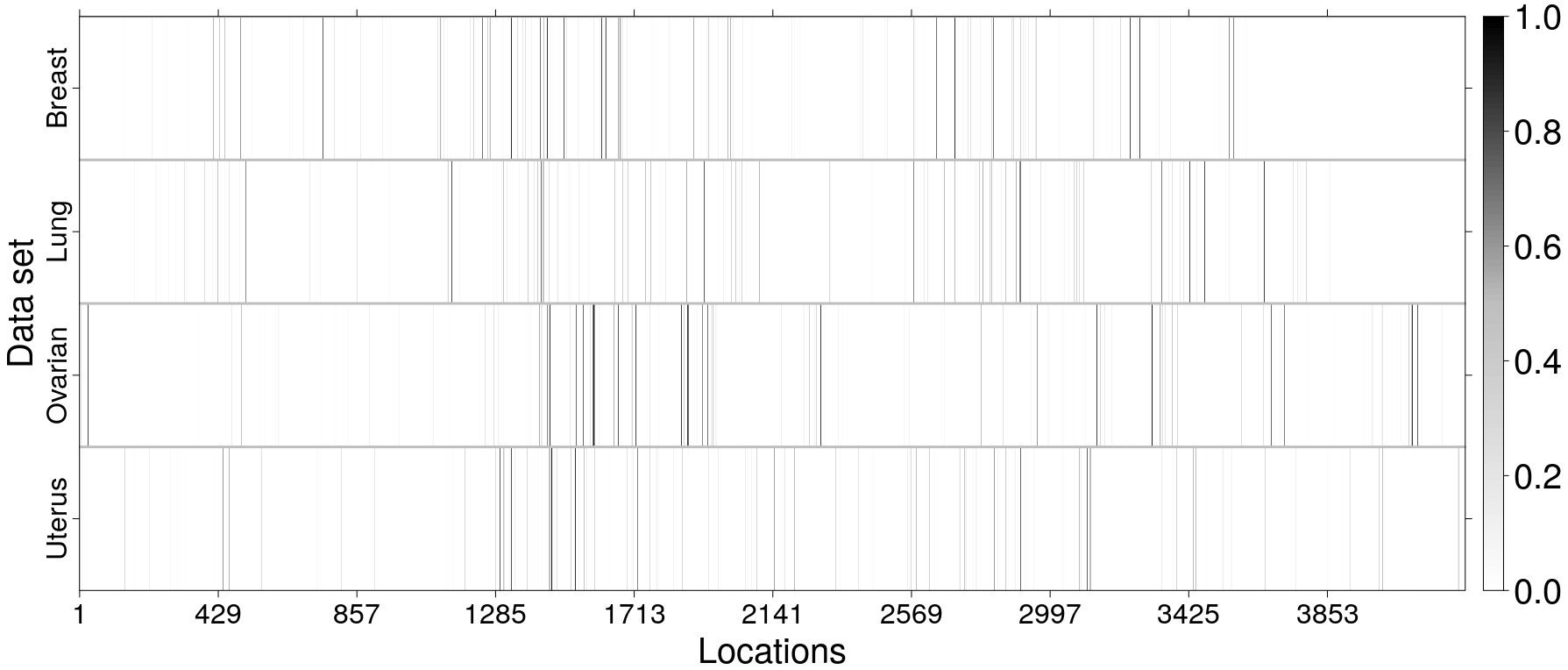} & \includegraphics[scale=0.16]{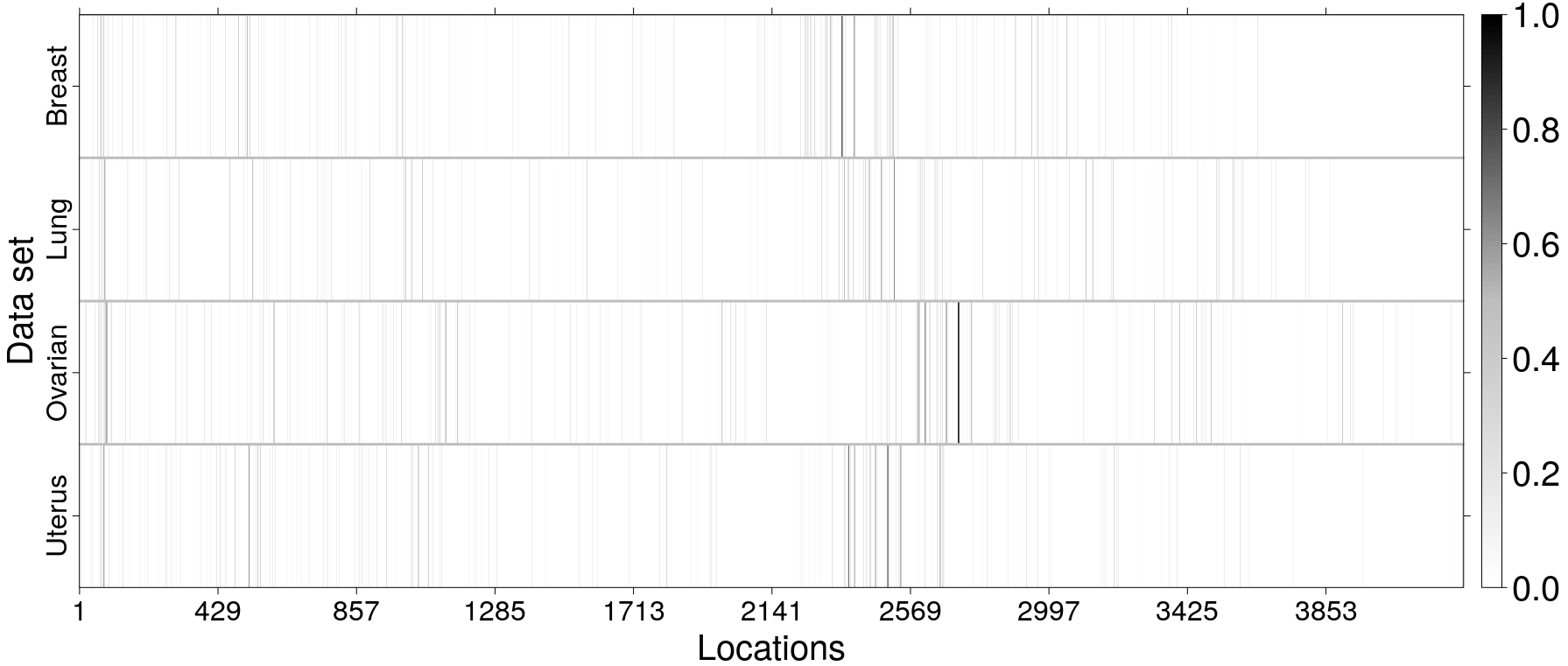}\\
  \end{array}
$$
\vspace{-10pt} \caption{Heat map image indicating, for each location, the posterior probability of belonging to a Gaussian component. Configuration of panels: lower Gaussian in column 1 (a, c), upper Gaussian in column 2 (b, d), all chromosomes in row 1 (a, b) and chromosome 1 in row 2 (c, d).}
\label{fig6}
\end{figure}

Figure \ref{fig6} presents grey scale heat maps to investigate the posterior probabilities of down and up-regulation in panels (a, c) and (b, d), respectively. Results for all $23$ chromosomes are shown in panels (a, b) and only chromosome 1 is explored in panels (c, d). Panel (d) shows that the Lung and Uterus cancers tend to provide coherent results in chromosome 1.

\begin{table}[h!]
\centering \small
\caption{Cross-comparison of data sets. Number of locations identified in the same Gaussian component (lower or upper) in both data sets, relative to the number of locations identified in that Gaussian component in at least one of the data sets. The results are presented as percentages.} \label{tab5}
\begin{tabular}{rrrrrrrrrr}
\hline
& \multicolumn{4}{c}{lower Gaussian} & & \multicolumn{4}{c}{upper Gaussian} \\
\cline{2-5} \cline{7-10}
& Breast & Lung & Ovarian & Uterus & & Breast & Lung & Ovarian & Uterus \\
\hline
Breast & 100 & 1.76 & 1.40 & 1.77 &  & 100 & 0 & 0 & 0 \\
Lung &  & 100 & 1.72 & 3.50 &  &   & 100 & 0 & 1.45 \\
Ovarian &  &  & 100 & 2.00 &  &  &  & 100 & 0 \\
Uterus &  &  &  & 100 &  &  &  &  & 100 \\
\hline
\end{tabular}
\end{table}

If we consider the strategy discussed in Section \ref{ssecclus} for cluster detection, i.e. finding consecutive genes with high probability ($> 0.5$) of down or up-regulation status, the following range of five locations are examples of under and overexpressed clusters, respectively: $6{,}238-6{,}242$ and $2{,}355-2{,}359$ (Breast), $4{,}355-4{,}359$ and $39{,}355-39{,}359$ (Lung), $10{,}679-10{,}683$ and $2{,}699-2{,}703$ (Ovarian) and $4{,}289-4{,}293$ and $37{,}364-37{,}368$ (Uterus). Different cluster sizes can be found when assuming this classification rule; the shortest case is a single observation surrounded by two internal Gaussian neighbours. The longest sequence, comparing all data sets, involves 9 locations (down-regulation) and 6 locations (up-regulation). 

\begin{figure}[!h]
$$
  \begin{array}{c}
   \includegraphics[scale=0.27]{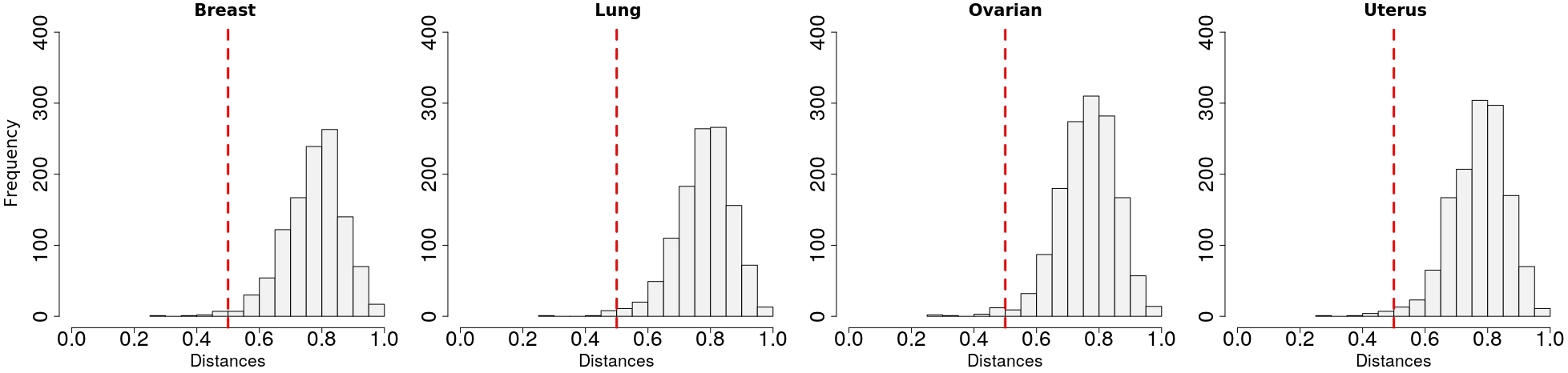} \\
   \includegraphics[scale=0.27]{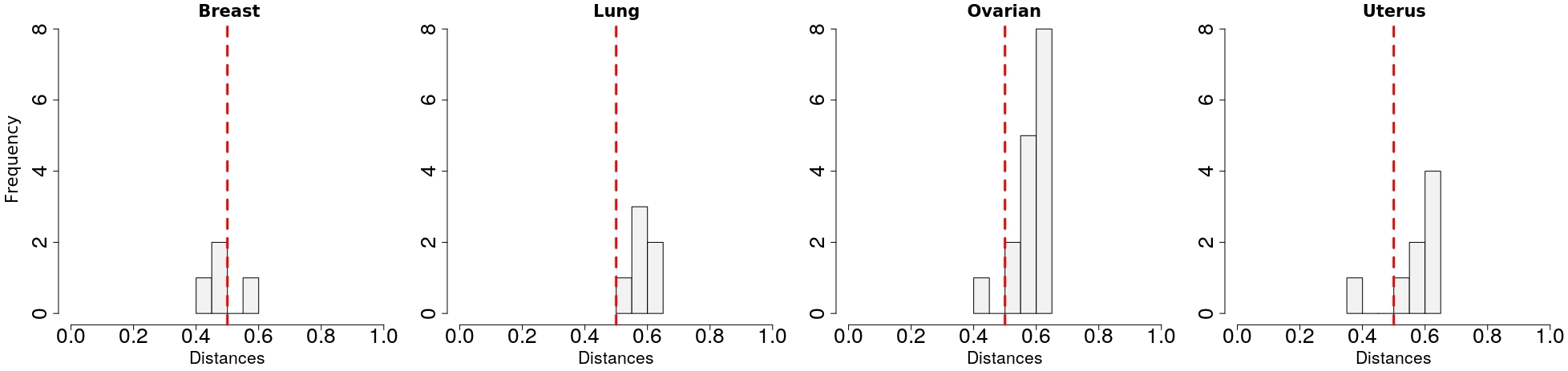} \\
  \end{array}
$$
\vspace{-10pt} \caption{Histogram of the distances between the location of a cluster, with a single observation, and its nearest internal Gaussian neighbours; down-regulation in row 1 and up-regulation in row 2.}
\label{fig7}
\end{figure}

The histograms in Figure \ref{fig7} show the distributions of all distances between an isolated under or overexpressed gene and its non-differentially expressed neighbours. Given that the upper Gaussian component accommodates less genes than the lower Gaussian, the histograms in row 2 are built using fewer distances. As it can be seen, in almost all cases the distribution is concentrated in the region above $0.5$. The only exception is the configuration ``isolated overexpressed gene in the breast cancer data''; however, the small bars here indicate very low frequency (breast data has the smallest genomic sequence) and they are clearly close to $0.5$. The overall behaviour exhibited in these graphs suggests that although the magnitude of an isolated expression is compatible with a down/up-regulated cluster, the model cannot use the neighbours to change this status, because they are too far away. These genes might belong to potential clusters, but a confirmation of this relies on their non-observed neighbours.

\begin{figure}[!h]
\centering
   \includegraphics[scale=0.27]{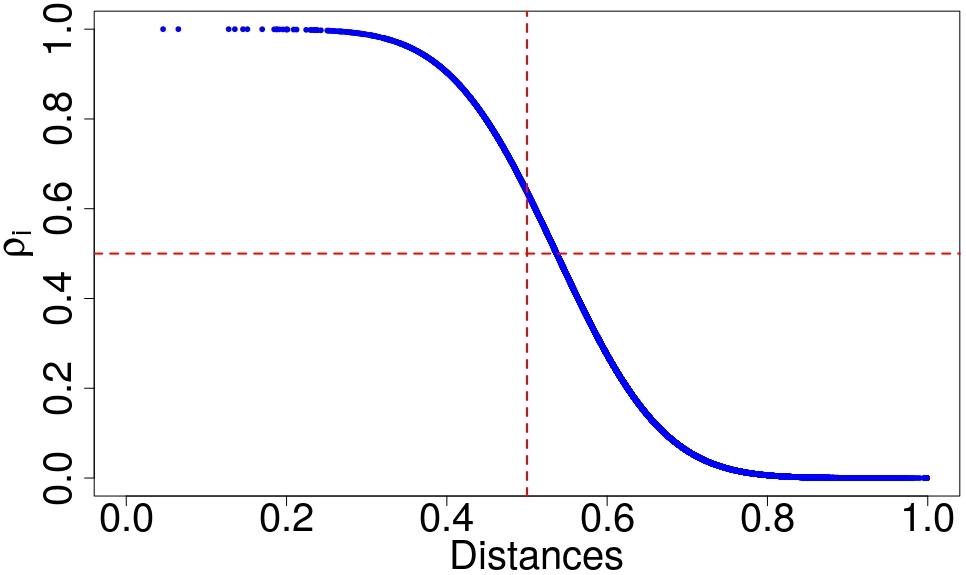} \\
\caption{Estimated relationship between the distances and $\rho_i$ (probability of having a Markov dependence). Result corresponding to the Breast data; the posterior means $5.108$ ($\beta_0$) and $-9.513$ ($\beta_1$) were used to draw the curve.}
\label{fig8}
\end{figure}

Figure \ref{fig8} presents the estimated relationship between $d_i$ and the probability $\rho_i$ of a Markov dependence at location $i$. The posterior means of ($\beta_0, \beta_1$) for each data set are, respectively: $5.108$ and $-9.513$ (Breast), $5.115$ and $-9.509$ (Lung), $5.149$ and $-9.486$ (Ovarian) and $5.121$ and $-9.504$ (Uterus). These are all similar, which is expected since we have assumed an informative prior for these coefficients. The decreasing rate of the curve is an important feature to explain the impact of $d_i$ on $\rho_i$; for example, the slow decay before distance $0.4$ indicates that the Markov dependence is strong for $d_i < 0.4$. In addition, the curve has a fast decay between distances $0.4$ and $0.6$, and it is almost $0$ for $d_i > 0.8$. This decreasing rate resembles the one observed in \cite{Mayrink17} due to our informative prior specification. 

%%%%%%%%%%%%%%%%%%%%%%
\section{Conclusions}

This paper proposes a statistical methodology to identify chromosomal regions associated with low/high gene expression measurements, which we call under/overexpressed or down/up-regulated regions, respectively, based on RNA-Seq data. Given the alignment of read counts to gene locations in the human chromosomes, the data is configured as an irregularly spaced sequence along the genome. Four data sets, representing breast, lung, ovarian and uterus cancer, are considered in the study; preprocessing and  normalising steps were applied to the raw read counts before the statistical analyses. The proposed methodology is based on the work from \cite{Mayrink17} with two main differences: RNA-Seq data is considered (instead of microarray) and both under and overexpressed chromosome regions are identified (not only the ``over'').

The proposed model is based on a mixture of Gaussians, such that the mixture components are supposed to classify the observations in three main categories; the lower Gaussian (smallest mean) accommodates the down-regulated expressions, the upper Gaussian (largest mean) incorporates the up-regulated expressions and all intermediate Gaussian components are related to the non-differentially expressed genes. The model takes advantage of the distances between gene locations to explain the Markov dependence between neighbours. This dependence is used to stochastically define the clusters of interest. The Bayesian inference is performed through an MCMC algorithm consisting of a Gibbs sampling.

The main results indicate that the model is able to recognise locations belonging to under and overexpressed regions. This identification is not arbitrary since the whole model structure, including prior specifications, has contributed to a selective judgment and clear discernment of the gene statuses through posterior probabilities. On average, only $2.02\%$ and $0.12\%$ of the genes reached a probability higher then $0.5$ to be considered (according to this classification rule) down and up-regulated cases, respectively. In a cross-comparison of cancer data sets and looking at the pattern exhibited by heat map graphs of probabilities, some few similarities can be detected suggesting the same differentially expressed location in two different cancers. The highest percentage of intersection is found in the Lung/Uterus comparison.

The cluster identification analysis presented in this paper is general and aims to provide a global picture of the down/up-regulated regions across the human genome. The proposed model can be used to study specific regions of interest in any chromosome, since it allows the calculation of the posterior probability that a particular set of probes form a down/up-regulated cluster as suggested in (\ref{clustprob}).

\vspace{10pt}

%%%%%%%%%%%%%%%%%%%%%%%%%%%%%%%%%%%%%%%%%%%%%%%%%%%%%%%%%%%
{\flushleft \sffamily \textbf{Acknowledgements}}
%%%%%%%%%%%%%%%%%%%%%%%%%%%%%%%%%%%%%%%%%%%%%%%%%%%%%%%%%%%
\vspace{5pt}

The authors would like to thank Funda\c{c}\~{a}o de Amparo \`{a} Pesquisa do Estado de Minas Gerais (FAPEMIG) for supporting this research.

\vspace{10pt}

%%%%%%%%%%%%%%%%%%%%%%%%%%%%%%%%%%%%%%%%%%%%%%%%%%%%%%%%%%%
\renewcommand{\thefigure}{A.\arabic{figure}} \setcounter{figure}{0}
\renewcommand{\theequation}{A.\arabic{equation}} \setcounter{equation}{0}
\renewcommand{\thetable}{A.\arabic{table}} \setcounter{table}{0}
{\flushleft \textbf{Appendix A: Joint and full conditional distributions} }
%%%%%%%%%%%%%%%%%%%%%%%%%%%%%%%%%%%%%%%%%%%%%%%%%%%%%%%%%%%
\vspace{10pt}

The joint density of $X$ and all the unknown components in the model is:
{\small
\begin{eqnarray}
\ds \pi(X,Z,W,V,q_0,Q,\psi,\beta) & = &\left[\prod_{i=1}^n \pi(X_i|Z_i,\psi) \, \pi(Z_i|Z_{i-1},W_i,q_0,Q) \, \pi(W_i|V_i) \, \pi(V_i|\beta) \right] \, \pi(q_0,Q,\psi,\beta),  \nonumber \\
    & = & \Bigg[ \prod_{i=1}^n \left[ \prod_{k=1}^{K} f_k(X_i|\psi)^{Z_{i,k}} (q_{0k}^{Z_{i,k}})^{1-W_i}(q_{k_{(i-1)}k}^{Z_{i,k}})^{W_i} \right] \nonumber \\
    &   & \times \; \left[ \mathds{1}(W_i = 1) \; \mathds{1}(V_i > 0) + \mathds{1}(W_i = 0) \; \mathds{1}(V_i \leq 0) \right] \nonumber \\
    &   & \times \; \phi(V_i-\beta'\vec{d_i}) \Bigg] \left[ \prod_{k=1}^{K} q_{0k}^{Z_{0,k}}q_{0k}^{r_{0k}-1} \right] \; \left[ \prod_{k_1=1}^{K} \prod_{k=1}^{K}q_{k_1 k}^{r_{k_1 k}-1} \right] \; \pi(\psi) \; \pi(\beta),
    \label{jdens}
\end{eqnarray}}
where $k_{(i-1)} = j$ if $Z_{i-1,j}=1$, \, and \, $\pi(\beta)=|\Sigma_0|^{-1/2}\phi_2[\Sigma_{0}^{-1/2}(\beta-\mu_0)]$. Notation: $\phi(.)$ and $\phi_2(.)$ are the densities of the uni- and bi-dimensional standard normal distributions, respectively.
{\small
\begin{eqnarray}
 \pi(\psi) & = & \ds \prod_{k=1}^{K} \; \pi_{\mbox{NIG}}(\mu_k,\sigma^2_k; m_k, v_k, s_{1k},s_{2k}) \; \mathds{1}(\mu_1<\ldots<\mu_K). \nonumber
\end{eqnarray}}

The full conditional distribution of $(q_0, Q, \psi, \beta)$ is:
\begin{eqnarray}
  (q_0|\cdot) &\sim& \mbox{Dir} \left[ r_{0} + \sum_{i=1}^n Z_{i} (1-W_i) \right], \label{fc1}\\
  (q_k|\cdot) &\sim& \mbox{Dir} \left[ r_{k} + \sum_{i=2}^n (Z_{i-1,k} W_i) Z_{i} \right], \label{fc2}\\
  (\beta|\cdot) &\sim& \mbox{N}_2(\mu_0^*,\Sigma_0^*),\label{fc3}
\end{eqnarray}
with $\Sigma_0^* = \left( \Sigma_0^{-1} + \sum_{i = 1}^n \vec{d}_i \vec{d_i}' \right)^{-1}$ and \; $\mu_0^* = \Sigma_0^* \left( \Sigma_0^{-1} \mu_0 + \sum_{i=1}^n V_i \vec{d}_i \right)$.

\begin{equation}
  (\mu_k|\cdot) \sim \mbox{N}(m_k^*,v_k^*) \quad \mbox{and} \quad (\sigma^2_k|\cdot) \sim \mbox{IG}(s_{1k}^*,s_{2k}^*),  \quad  \mbox{for} \; k = 1, \ldots, K, \label{fc4}
\end{equation}
where \; $v_k^* = \frac{v_k \sigma^2_k}{1 + v_k \sum_{i=1}^n Z_{i,k}}$, \; $m_k^* = v_k^* \left( \frac{m_k + v_k \sum_{i=1}^n Z_{i,k} X_i}{v \sigma^2_k} \right)$, \;
$s_{1k}^* = s_{1k} + \frac{1}{2} \left( \sum_{i=1}^n Z_{i,k} \right)$ \; and \\
$s_{2k}^* = s_{2k} + \frac{1}{2} \left[ \frac{m_k^2}{v_k} + \sum_{i=1}^n Z_{i,k} X_i^2 -\frac{v_k}{1+v_k \sum_{i=1}^n Z_{i,k}} \left( \frac{m_k}{v_k} + \sum_{i=1}^n Z_{i,k} X_i \right)^2 \right]$. \\

The sampling step of $(Z,W)$ is  \vspace{-3pt}
\begin{eqnarray}
 Z_1  & \sim & \mbox{Mult}(1,q_{0}^*), \nonumber \\
 (W_i|Z_{i-1,j}=1,\cdot) & \sim & \mbox{Ber}(1,p_{(i,j)}^*), \;\; i = 2, \ldots, n, \nonumber \\
 (Z_i|W_i=l,Z_{i-1,j}=1,\cdot) & \sim & \mbox{Mult}(1,q_{(i,j,l)}^*), \;\; i = 2, \ldots, n, \nonumber
\end{eqnarray}
where \vspace{-3pt}
\begin{eqnarray} \label{fsampling}
 q_{0k}^* &=& c_{1,k}q_{0k}/a_1, \hspace{2.3cm} p_{(i,j)}^* = b_{i,j}\Phi_{i}^+/c_{i,j}, \\
 q_{(i,j,0)k}^* &=& c_{i+1,k}f_k(X_i|\psi)q_{0k}/a_i, \quad q_{(i,j,1)k}^*  = c_{i+1,k}f_k(X_i|\psi)q_{j,k}/b_{i,j}. \nonumber
\end{eqnarray}
Consider $c_{n+1,j}=1$, \, $\forall j$, \, and:
\begin{eqnarray} \label{filtering}
  a_n & = & \sum_{k=1}^{K}f_k(X_n|\psi)q_{0k}, \quad b_{n,j} = \sum_{k=1}^{K}f_k(X_n|\psi)q_{j,k}, \quad c_{n,j}=b_{n,j}\Phi_{n}^++a_n\Phi_{n}^- , \\
  a_{i} & = & \sum_{k=1}^{K}c_{i+1,k}f_k(X_i|\psi)q_{0k}, \quad b_{i,j} = \sum_{k=1}^{K}c_{i+1,k}f_k(X_i|\psi)q_{j,k}, \quad c_{i,j} = b_{i,j}\Phi_{i}^++a_i\Phi_{i}^-, \nonumber \\
  & & \mbox{for }i=n-1,\ldots,2, \nonumber \\
  a_{1} & = & \sum_{k=1}^{K}c_{1,k}q_{0k}. \nonumber
\end{eqnarray}
First, the calculations are performed recursively, starting from $n$ and moving backwards in the filtering part given in (\ref{filtering}); here, $b_i$ and $c_i$ are $K$-vectors and $a_i$ is a scalar. Next, the probabilities shown in (\ref{fsampling}) are calculated and  $(Z,W)$ are sampled.

\vspace{5pt}

%%%%%%%%%%%%%%%%%%%%%%%%%%%%%%%%%%%%%%%%%%%%%%%%%%%%%%%%%%%
\renewcommand{\thefigure}{B.\arabic{figure}} \setcounter{figure}{0}
\renewcommand{\theequation}{B.\arabic{equation}} \setcounter{equation}{0}
\renewcommand{\thetable}{B.\arabic{table}} \setcounter{table}{0}
{\flushleft \textbf{Appendix B: Sensitivity analysis} }
%%%%%%%%%%%%%%%%%%%%%%%%%%%%%%%%%%%%%%%%%%%%%%%%%%%%%%%%%%%
\vspace{10pt}

\begin{figure}[!h]
\centering
 $$
 \begin{array}{c}
 \includegraphics[scale=0.27]{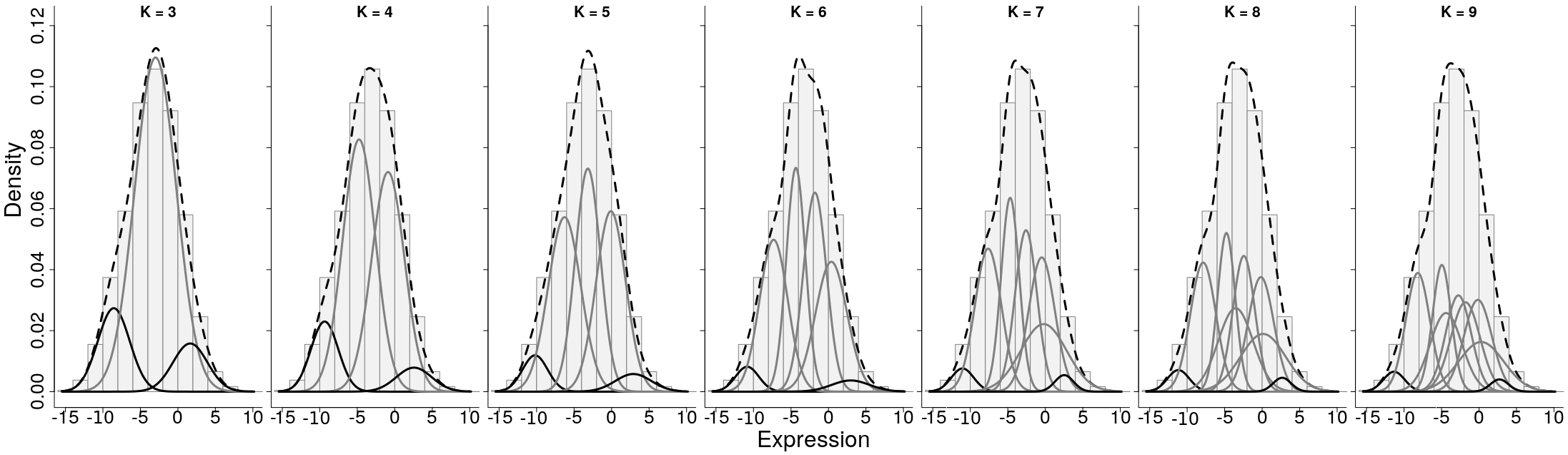} \\
 \includegraphics[scale=0.27]{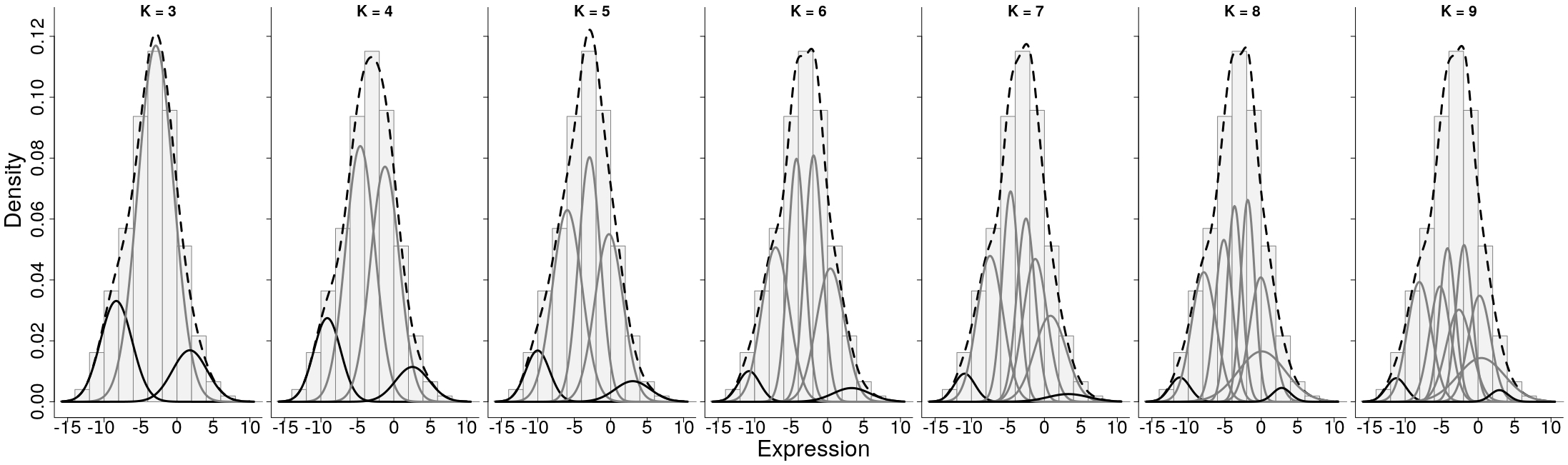} \\
 \end{array}
 $$
\vspace{-0.7cm}
\caption{Histogram of all expressions (Breast in row 1, Ovarian in row 2) overlaid by the estimated mixture density (dashed curve) and its Gaussian components (grey and black curves) for seven different choices of $K$.} \label{figB1}
\end{figure}

Figure \ref{figB1} and Table \ref{tabB1} report the sensitivity of the results to the choice of $K$ (number of Gaussian components). Here, the model is fitted assuming seven configurations: $K = 3$, $4$, $5$, $6$, $7$, $8$ and $9$. The Dirichlet priors for $q_0$ and $Q$ are specified so that the degree of information in the prior (sum of the hyperparameters of the Dirichlet) is the same across all configurations; see the description in Section \ref{secresult}. As expected, the results for $K = 3$ indicates lower/upper Gaussian components with high variance and accommodating too many expressions. The weights of the lower/upper Gaussian components decreases as $K$ increases. In terms of variance, the two smallest values are observed for large $K$ in both cases. Table \ref{tabB1} also shows that the variance of the mixture involving only the internal normals is increasing with $K$. Note that the results become quite similar for $K = 7$ and $8$ (Breast) and $K = 8$ and $9$ (Ovarian). Recall that the Ovarian data set is larger (6{,}541 more aligned genes) than the Breast data; this might explain the differences observed between them. In conclusion, it seems reasonable to choose the model with $K = 8$ (results shown in Section \ref{secresult}).

\begin{table}[!h]
\centering \scriptsize
\setlength\tabcolsep{2.8pt}
\caption{Posterior estimates (for each size $K$, Ovarian data) of the weight, $\mu_k$ and $\sigma^2_k$ related to the lower and upper Gaussian components and the mean (expec.)  and variance (var.) of the (normalised)
mixture of internal normals (without the lower and upper components); standard errors in parentheses.}
\label{tabB1}
\begin{tabular}{c|ccc|ccc|cc}
\hline
  $K$  & weight $1$ & $\mu_1$ & $\sigma^2_1$ & weight $K$ & $\mu_K$ & $\sigma^2_K$ & expec. & var. \\
\hline
$3$ & 0.180 (0.007) &   -8.331 (0.088) & 4.705 (0.168) & 0.101 (0.005) & 1.760 (0.093) & 5.724 (0.209) & -2.914 (0.041) &   6.008 (0.130) \\
$4$ & 0.128 (0.008) &   -9.130 (0.114) & 3.462 (0.159) & 0.066 (0.005) & 2.551 (0.124) & 5.318 (0.268) & -2.995 (0.058) &   6.851 (0.157) \\
$5$ & 0.069 (0.006) & -10.020 (0.110) & 2.704 (0.137) & 0.041 (0.004) & 3.022 (0.188) & 5.877 (0.432) & -3.197 (0.048) &   8.765 (0.187) \\
$6$ & 0.037 (0.004) & -10.763 (0.124) & 2.201 (0.157) & 0.029 (0.003) & 3.296 (0.233) & 6.483 (0.576) & -3.327 (0.031) & 10.071 (0.167) \\
$7$ & 0.033 (0.004) & -10.941 (0.134) & 2.057 (0.163) & 0.018 (0.003) & 3.283 (0.366) & 7.803 (1.085) & -3.276 (0.032) & 10.638 (0.179) \\
$8$ & 0.028 (0.003) & -11.157 (0.130) & 1.914 (0.158) & 0.014 (0.002) & 2.802 (0.197) & 1.506 (0.339) & -3.280 (0.027) & 11.174 (0.152) \\
$9$ & 0.026 (0.003) & -11.228 (0.142) & 1.855 (0.168) & 0.012 (0.002) & 2.913 (0.258) & 1.482 (0.391) & -3.276 (0.027) & 11.265 (0.155) \\
\hline
\end{tabular}
\end{table}

\begin{figure}[!h]
\centering
   \includegraphics[scale=0.25]{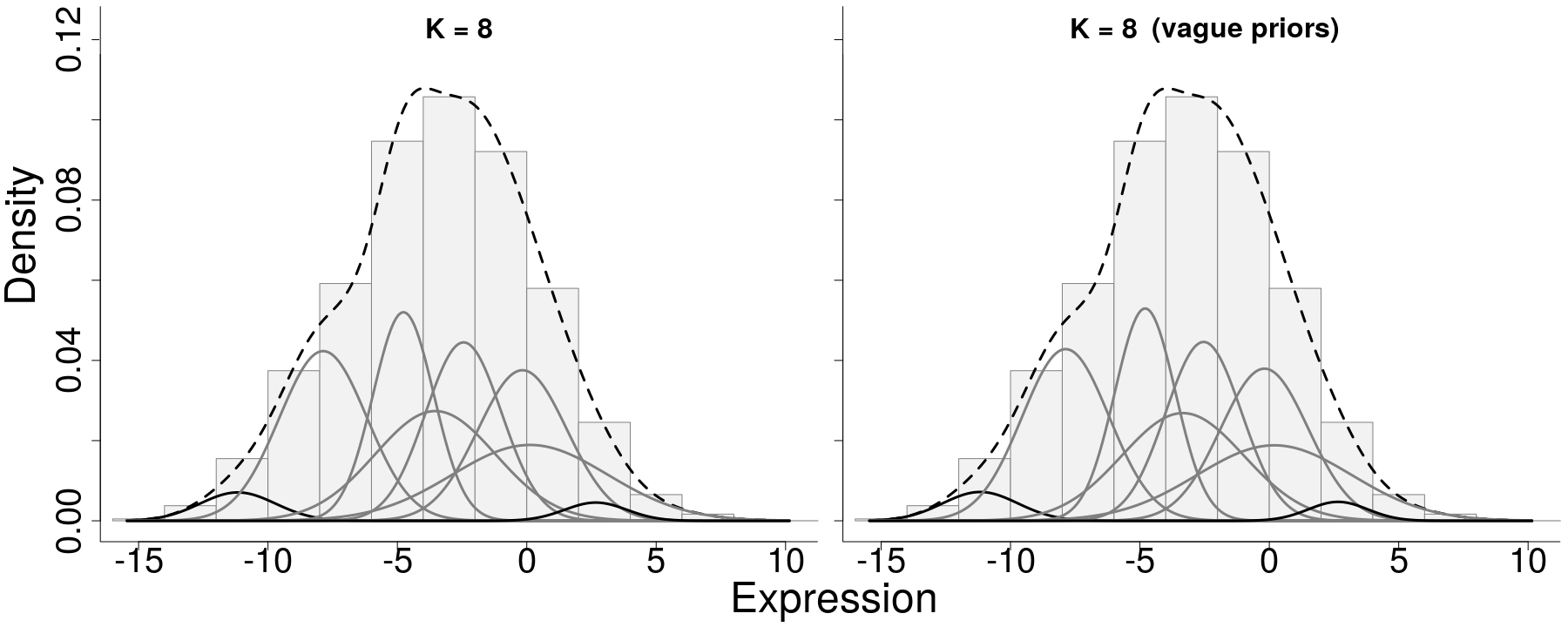}
\caption{Comparison of prior specifications (Breast data). Histogram of all expressions overlaid by the estimated mixture density (dashed curve) and its components (grey and black curves).}
\label{figB2}
\end{figure}

The next results are related to a prior sensitive analysis developed for the parameters of the Gaussian components in the mixture with $K=8$. Two different specifications are studied here, the first one has the same configuration explored in Section \ref{secresult}. Consider $(\mu_k, \sigma^2_k) \sim \mbox{NIG}(m_k,\;10,\; 2.1,\;1.1)$ with $m_k = -10$, $-7.14$, $-4.29$, $-1.43$, $1.43$, $4.29$, $7.14$ and $10$, for $k = 1, \ldots, 8$, respectively. The second specification assumes higher variability by doubling the original standard deviation, i.e., $(\mu_k, \sigma^2_k) \sim \mbox{NIG}(m_k,\;40,\; 2.025,\;1.025)$; the values of $m_k$ and $E(\sigma^2_k)$ are not changed with respect to the first option. Figure \ref{figB2} and Table \ref{tabB2} show the results; they are visually the same and this indicates robustness to the prior uncertainty of these parameters.

\newpage

\begin{table}[!h]
\centering \scriptsize
\setlength\tabcolsep{2.8pt}
\caption{Comparison of prior specifications (Breast data). Posterior estimates of: the weights, $\mu_k$, $\sigma^2_k$, for $k = 1$ and $8$ (lower and upper Gaussian components), mean (expec.) and variance (var.) of the normalised mixture of internal normals; standard errors in parentheses.} \label{tabB2}
\begin{tabular}{r|ccc|ccc|cc}
\hline
Prior & weight $1$ & $\mu_1$ & $\sigma^2_1$ &  weight $8$ & $\mu_8$ & $\sigma^2_8$ & expec. & var. \\
\hline
Original & 0.026 (0.003) & -11.168 (0.162) & 2.125 (0.200) & 0.014 (0.002) & 2.666 (0.312) & 1.436 (0.350) & -3.247 (0.029) & 11.666 (0.158) \\
Vague    & 0.026 (0.003) & -11.170 (0.155) & 2.111 (0.190) & 0.014 (0.002) & 2.657 (0.286) & 1.312 (0.351) & -3.244 (0.028) & 11.653 (0.153) \\
\hline
\end{tabular}
\end{table}

\vspace{5pt}

%%%%%%%%%%%%%%%%%%%%%%%%%%%%%%%%%%%%%%%%%%%%%%%%%%%%%%%%%%%
\renewcommand{\thefigure}{C.\arabic{figure}} \setcounter{figure}{0}
\renewcommand{\theequation}{C.\arabic{equation}} \setcounter{equation}{0}
\renewcommand{\thetable}{C.\arabic{table}} \setcounter{table}{0}
{\flushleft \textbf{Appendix C: MCMC diagnostics} }
%%%%%%%%%%%%%%%%%%%%%%%%%%%%%%%%%%%%%%%%%%%%%%%%%%%%%%%%%%%
\vspace{10pt}

\begin{figure}[!h]
\centering
   \includegraphics[scale=0.27]{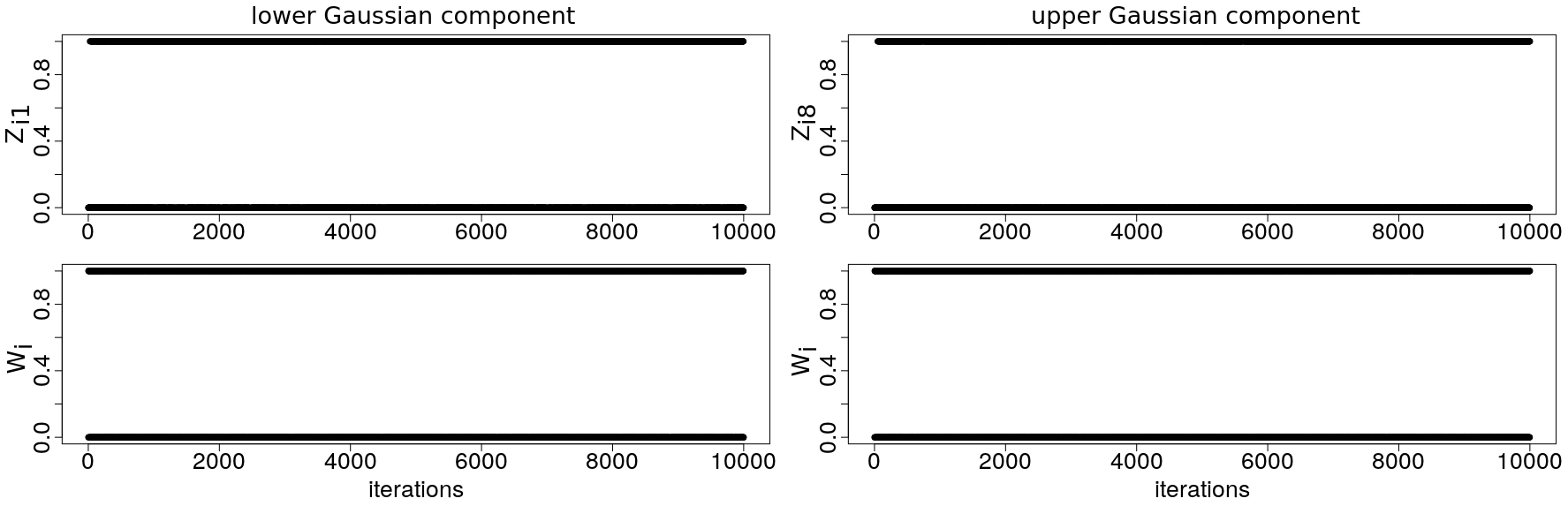}
\caption{Trace plots of: $Z_{i1}$ and $W_i$ (from location $i = 1{,}379$) and $Z_{i8}$ and $W_i$ (from location $i = 24{,}714$). These are locations with posterior mean near $0.5$ (Breast data).}
\label{figC1}
\end{figure}

\begin{figure}[!h]
\centering
   \includegraphics[scale=0.27]{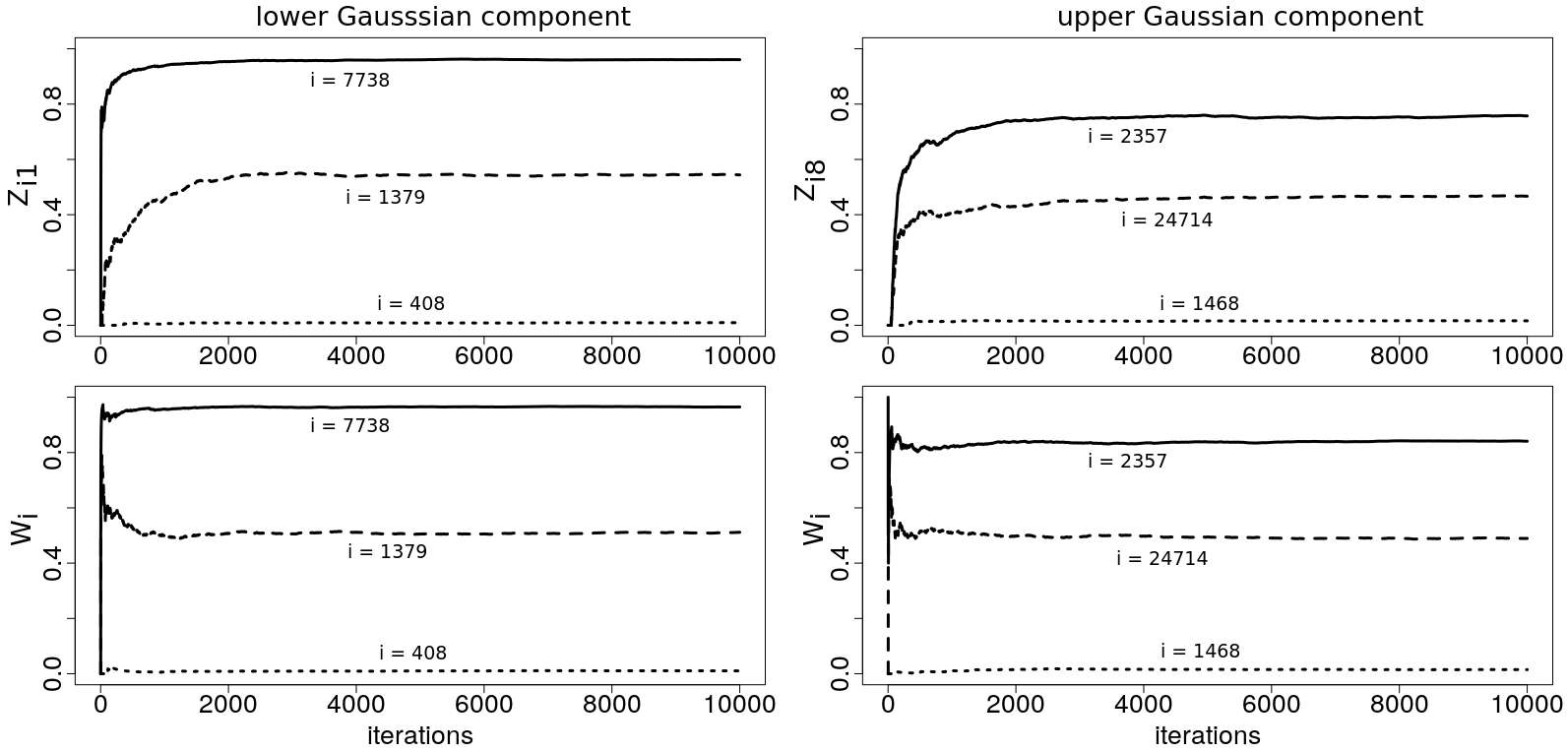}
\caption{Evolution of the ergodic average for three pairs of $(Z_{i1},W_i)$ and three pairs of $(Z_{i8},W_i)$ along the MCMC chain (Breast data).}
\label{figC2}
\end{figure}

This section presents some results to show that the MCMC designed for our gene expression analysis has good convergence properties. Two aspects explaining this behaviour are: the chosen blocking scheme for the algorithm and the fact that all parameters, including ($Z, W$), can be directly sampled from their full conditional distributions. The graphs in Figures \ref{figC1} and \ref{figC2} explore some MCMC chains for the block $(Z, W)$. Here, the trace plots indicate good mixing properties and the trajectories of the ergodic averages suggest fast convergence.

\newpage

%%%%%%%%%%%%%%%%%%%%%%%%%%%%%%%%%%%%%%%%%%%%%%%%%%%%%%%%%%%
%%%%%%%%%%%%%%%%%%%%%%%%%%%%%%%%%%%%%%%%%%%%%%%%%%%%%%%%%%%
%% Bibliography BibTeX.
{\bibliographystyle{Chicago}
\setlength{\bibsep}{0.3pt}
\small \bibliography{references}}

%%%%%%%%%%%%%%%%%%%%
\end{document}